\documentclass[11pt,a4paper]{article}
\usepackage{graphicx}
\usepackage{epstopdf}
\usepackage{amsmath,amssymb,amsbsy,color}
\usepackage{grffile} 
\usepackage{multirow}
\usepackage{hyperref}
\usepackage[ruled,vlined,noline]{algorithm2e}
\usepackage{algorithmic}
\usepackage[numbers,sort&compress]{natbib}
\begin{document}
\begin{flushright}
  \small{
    UTHEP-714
    UTCCS-P-110
  }
\end{flushright}
\vspace{2mm}

\begin{center}
{\Large\bf Finite-volume correction on the hadronic vacuum polarization contribution to muon g-2 in lattice QCD}
\end{center}
\vspace{5mm}
\begin{center}
  Taku~Izubuchi$^{(a,b)}$,
  Yoshinobu~Kuramashi$^{(c,d)}$,
  Christoph~Lehner$^{(b)}$,
  Eigo~Shintani$^{(d)}$\\
  (PACS collaboration)
\\[4mm]
{\small\it
  $^a$RIKEN-BNL Research Center, Brookhaven National Laboratory, Upton, NY 11973, USA}
\\
{\small\it
  $^b$Physics Department, Brookhaven National Laboratory, Upton, NY 11973, USA}
\\
{\small\it
  $^c$Center for Computational Sciences, University of Tsukuba, Tsukuba, Ibaraki 305-8577, Japan}
\\
{\small\it
  $^d$RIKEN Center for Computational Science, Kobe, Hyogo 650-0047, Japan}
\\[10mm]
\end{center}

\begin{abstract}
  We study the finite-volume correction on the hadronic vacuum polarization contribution to the muon g-2 ($a_\mu^{\rm hvp}$) in lattice QCD at (near) physical pion mass using two different volumes: $(5.4~{\rm fm})^4$ and $(8.1~{\rm fm})^4$.  We use an optimized AMA technique for noise reduction on $N_f=2+1$ PACS gauge configurations with stout-smeared clover-Wilson fermion action and Iwasaki gauge action at a single lattice cut-off $a^{-1}=2.33$ GeV.  The calculation is performed for the quark-connected light-quark contribution in the isospin symmetric limit.  We take into account the effects of backward state propagation by extending a temporal boundary condition. In addition we study a quark-mass correction to tune to the exactly same physical pion mass on different volume and compare those correction with chiral perturbation. We find $10(26)\times10^{-10}$ difference for light quark $a_\mu^{\rm hvp}$ between $(5.4~{\rm fm})^4$ and $(8.1~{\rm fm})^4$ lattice in 146 MeV pion. 
\end{abstract}

\section{Introduction}

The muon anomalous magnetic moment (g-2) is an essential observable for a rigorous test of the standard model (SM) of particle physics. The experimental value of the muon g-2, $a_\mu^{\rm E821}=11~659~209.1(5.4)(3.3)\times 10^{-10}$, has been measured more than a decade ago at BNL in the E821 experiment~\cite{Bennett:2004pv,Bennett:2006fi} and currently exhibits a 3-4 $\sigma$ tension with the SM theory prediction.  Within the theory prediction the QED corrections are now known to 5-loop order~\cite{Aoyama:2012wj} surpassing the precision of $a_\mu^{\rm (E821)}$ by two orders of magnitude.  The electroweak interaction contribution involving $W^\pm$, $Z$, and the Higgs is known one order-of-magnitude more precisely than $a_\mu^{\rm E821}$.  The theory uncertainty is currently dominated by the leading-order QCD contribution, {\it i.e.}, the hadronic vacuum polarization (HVP) contribution.  This contribution is typically extracted from $e^+e^-\rightarrow$hadron \cite{Babusci:2012rp,Babusci:2012ft,Lees:2012cj,Ablikim:2015orh}  or $\tau\rightarrow$hadron decays \cite{Ackerstaff:1998yj,Anderson:1999ui,Schael:2005am,Fujikawa:2008ma} using a dispersion relation \cite{Davier:2010nc,Hagiwara:2011af,Davier:2017zfy,Keshavarzi:2018mgv} and currently has an uncertainty similar to the error of $a_\mu^{\rm E821}$.  A comparable uncertainty comes from the hadronic light-by-light contribution whose model-dependence is still under scrutiny~\cite{Jegerlehner:2009ry}. To resolve the discrepancy between experiment and the SM calculation we need to reduce the uncertainties of both QCD contributions and the experiment. The upcoming experiments at Fermilab~\cite{Flay:2016vuw} and J-PARC~\cite{Shimomura:2015aza} aim for a four-fold improvement over $a_\mu^{\rm E821}$ in the near future which makes a similar precision improvement of the HVP contribution of timely interest.

The current estimate of the leading order of hadronic contribution (HLO) to muon g-2 is $a_\mu^{\rm HLO}=(693.3(2.5)\times 10^{-10}$ for $e^+e^-$ collision data and $688.9(3.5)\times 10^{-10}$ for $\tau$-decay data quoted from \cite{Keshavarzi:2018mgv,Olive:2016xmw,Jegerlehner:2015stw}. The determination from $e^+e^-\rightarrow$hadrons($\gamma$) cross sections \cite{Gourdin:1969dm} as a function of center-of-mass energy includes the QED effect in the hadron vertex and radiative correction of the final state~\cite{Olive:2016xmw}.  On the other hand, lattice QCD, which is a rigorous computation from the first principle of QCD, is able to provide the pure QCD contribution to $a_\mu^{\rm hvp}$ for the whole energy region, and its calculation is completely independent from the dispersive approach. Therefore a high-precision lattice QCD result is required for a cross-check of and potential improvement over the dispersive value. 

Recently the lattice QCD community has made significant progress to improve the precision of $a_\mu^{\rm hvp}$. Starting from quenched QCD calculations~\cite{Blum:2002ii,Gockeler:2003cw}, $N_f=2$ or 2+1 QCD calculations using various lattice fermion formulation and lattice parameters have been reported by several groups~\cite{Feng:2011zk,Boyle:2011hu,DellaMorte:2011aa,Burger:2013jya,Chakraborty:2014mwa,Chakraborty:2016mwy,Borsanyi:2016lpl,Borsanyi:2017zdw,DellaMorte:2017dyu} and recent lattice QCD calculation with perturbative QED correction at physical pion mass is now available \cite{Blum:2018mom}.  Currently, however, the precision of pure lattice calculations is about 5 times lower than of the dispersive approach.  One major source of uncertainty in a lattice QCD evaluation of $a_\mu^{\rm hvp}$ is the statistical noise of the Monte-Carlo method. In addition finite-volume (FV) correction for lattice size $L\sim5$--6 fm have been expected to be significant (see {\it e.g.} \cite{Blum:2018mom}) and it then has been so-far only treated by an effective-field theory \cite{Aubin:2015rzx,Bijnens:2017esv}.  An estimate of the FV correction with the pure lattice calculation is therefore highly desired to control systematic error in a lattice calculation for the desired precision.

In this paper, to study FV correction to $a_\mu^{\rm hvp}$ in purely lattice QCD, we compare the connected HVP diagram between two volumes, $L= 5.4$ and 8.1 fm, at nearly physical pion ($m_\pi\simeq 0.14$ GeV), which are corresponding to two variations of $m_\pi L= 3.8$ and 5.8. Since the statistical noise of the infrared region should be significantly reduced to perform a rigorous test of FV effect, we utilize a highly-optimized AMA technique reported in \cite{vonHippel:2016wid} which can further improve the performance rather than the original proposals~\cite{Blum:2012uh,Blum:2012my,Shintani:2014vja}. Our study also provides a test of usage of an effective-field theory for FV correction as used in \cite{Chakraborty:2016mwy,Borsanyi:2017zdw,Blum:2018mom} and cross-check from pure lattice calculation. 

This paper is organized as the followings. In section~\ref{sec:backgroud}, we introduce the method to compute $a_\mu^{\rm hvp}$. Section~\ref{sec:lattice_param} shows our setup for the numerical computation, and in section~\ref{sec:result} we present numerical results on two different volumes. In section~\ref{sec:mass_correction}, we discuss on mass correction and the FV effect obtained by our numerical study. Finally, in section~\ref{sec:summary}, we summarize this paper and discuss future extensions.

\section{Lattice computation of $a_\mu^{\rm hvp}$}\label{sec:backgroud}

Since the lattice QCD calculation is defined in Euclidean space-time, a conventional representation of $a_\mu^{\rm hvp}$ is as the integral of vacuum polarization function (VPF) $\Pi(Q)$ with respect to Euclidean momentum squared $Q^2$ from zero to infinity,
\begin{eqnarray}
  &&a_\mu^{\rm hvp} = \Big(\frac{\alpha_e}{\pi}\Big)^2\int^\infty_0 dQ^2 K_E(Q^2)\hat\Pi(Q^2), \quad
  \hat\Pi(Q^2) \equiv \Pi(Q^2)-\Pi(0),\label{eq:a_mu_infinite}\\
  &&K_E(s) = \frac{1}{m_\mu^2}\hat sZ^3(\hat s)\frac{1-\hat s Z(\hat s)}{1+\hat sZ^2(\hat s)},\\
  && Z(\hat s) = -\frac{\hat s-\sqrt{\hat s^2+4\hat s}}{2\hat s},\quad \hat s = \frac{s}{m_\mu^2},
\end{eqnarray}
which can be derived by analytic continuation from the original representation using time-like momentum $q^2$($=-Q^2$)~\cite{LAUTRUP1972193,Blum:2002ii}.  $K_E$ is a known QED kernel resulting from the needed one-loop computation and $\alpha_e$ is a fine structure constant $\alpha_e=1/137.03599914$. $\hat\Pi$ denotes the subtracted VPF at $Q^2=0$.  For the lattice computation of $a_\mu^{\rm hvp}$, due to the non-zero lattice spacing and finite volume, we convert its integral to a finite momentum sum.  In this paper we deal with the integral in Eq.~(\ref{eq:a_mu_infinite}) as the coordinate space-time summation of vector-vector current correlators on the lattice, in the so-called ``time-momentum representation'' (TMR) \cite{Bernecker:2011gh}. 

We use the vector-vector current correlator at zero momentum in spatial direction $i$, 
\begin{eqnarray}
  C(t) = \frac 1 3\sum_{i=1}^3\int d^3\vec x\langle V_i^{\rm cv}(\vec x,t)V_i^{\rm loc}(0)\rangle
\end{eqnarray}
with local lattice current
\begin{equation}
  V^{\rm loc}_\mu=Z_V\bar q(x)\gamma_\mu q(x)
\end{equation}
and Z factor $Z_V=0.95153(76)(1487)$, evaluated by the Schr\"odinger functional method \cite{Ishikawa:2015fzw}. We also use the conserved current 
\begin{eqnarray}
  V_\mu^{\rm cv}(x) = \frac{1}{2}\Big[ \bar q(x+a\hat\mu)(1+\gamma_\mu)U_\mu^{\dag}(x)q(x)
    - \bar q(x)(1-\gamma_\mu)U_\mu(x)q(x+a\hat\mu)\Big],
\end{eqnarray}
i.e., the point-split current that satisfies the Ward-Takahashi identity as also used in \cite{Shintani:2010ph,Boyle:2011hu,DellaMorte:2017dyu}. In the TMR, $a_\mu^{\rm hvp}$ in Eq.~(\ref{eq:a_mu_infinite}) can be also represented as 
\begin{eqnarray}
  a_\mu^{\rm hvp} &=& 4\alpha_e^2m_\mu\int^{\infty}_0 dt t^3 C(t)\tilde K(t)\label{eq:amu_TMR},\\
  \tilde K(t) &=& \frac{2}{m_\mu t^3}\int^{\infty}_0\frac{d\omega}{\omega} K_E(\omega^2)
  [\omega^2t^2-4\sin^2(\omega t/2)],
  \label{eq:k_tilde}
\end{eqnarray}
as shown in \cite{Bernecker:2011gh}.  On the lattice, the above becomes the summation of discretized $C(t)$ multiplied with $\tilde K(t)$ up to a half length of lattice temporal extension\footnote{In periodic or anti-periodic boundary conditions, BSP significantly alters the $C(t)$ at $t\sim N_t/2a$, which is one of the FV (or finite temporal extension) effects. We numerically study this in section~\ref{sec:result}.}. Setting the truncation bound of the sum to $t_{\rm cut}<N_ta/2$ in an integral of Eq.~\ref{eq:amu_TMR}, the lattice representation is
\begin{eqnarray}
  [a_\mu^{\rm hvp}]_{\rm lat}(t_{\rm cut}) &=& \frac1 2\sum_{t/a=0}^{t_{\rm cut}/a-1} \Big[C(t)W_t(t) + C(t+a)W_t(t+a)\Big],\label{eq:amu_TMR_lat}\\
  W_t(t) &=& 8\alpha_e^2\int^\infty_0\frac{d\omega}{\omega}K_E(\omega^2)\Big[\omega^2t^2-4\sin^2(\omega t/2)\Big],
\end{eqnarray}
in which the expression of $\tilde K$ in Eq.~(\ref{eq:k_tilde}) is substituted and the trapezoidal formula is used for numerical integral. 

We also note that the representation of Eq.~(\ref{eq:amu_TMR_lat}) is not unique for finite lattice spacing. For example, if we use the sin functional form of lattice momentum, $\tilde Q=2a^{-1}\sin(Q_\mu/2a)$, such a representation is changed to 
\begin{eqnarray}
  [\tilde a_\mu^{\rm hvp}]_{\rm lat}(t_{\rm cut}) &=& \frac 1 2\sum_{t/a=0}^{t_{\rm cut}/a-1} \Big[C(t)\tilde W(t)+C(t+a)\tilde W(t+a)\Big],\label{eq:amu_TMR_lat2}\\  
  \tilde W(t) &=& 8\alpha_e^2\int^\infty_0\frac{\omega d\omega}{\tilde\omega^2}K_E(\omega^2)\Big[\tilde\omega^2t^2-4\sin^2(\omega t/2)\Big],
\end{eqnarray}
where we use $\tilde\omega=2a^{-1}\sin(aQ/2)$. The trivial difference between Eq.~(\ref{eq:amu_TMR_lat}) and Eq.~(\ref{eq:amu_TMR_lat2}) is at $t=a$, in which the integrand of $[\tilde a_\mu^{\rm hvp}]_{\rm lat}$ is zero, besides that of $[a_\mu^{\rm hvp}]_{\rm lat}$ is non-zero. The difference between $[a_\mu^{\rm hvp}]_{\rm lat}$ and $[\tilde a_\mu^{\rm hvp}]_{\rm lat}$ can be used as a simple estimate of lattice artifacts (see Section~\ref{sec:summary}).

\newcommand{\ticm}[1]{{\color{red} [#1]}}

\subsection{Strategy to measure finite volume effect}
In this paper, we numerically estimate the finite volume (FV) correction in the TMR directly at physical pion mass by comparing two volumes at the same cut-off scale. This allows us to remove uncertainties due to the chiral extrapolation ansatz~\cite{Golterman:2017njs}, which becomes the large contribution around physical pion extrapolated from unphysically heavy mass~\cite{DellaMorte:2017dyu}. As we will see, pion masses on the two volumes are both very close to the physical mass, but there is a small difference between the two masses \cite{pacs2017}. To clearly separate the FV effect from the effects from using a slightly different pion mass, we correct this small mass difference by adjusting the valence light quark mass as well as the sea quark mass via the reweighting technique \cite{Hasenfratz:2008fg}.

One practical issue in using the TMR is that in order to evaluate the t-integral one needs to precisely evaluate vector-vector current correlators at a large distance before the integrand at infrared regime is negligibly small.  This is a concern since the lattice data is limited to $|t| < N_ta/2$ and the statistical noise grows exponentially with time. One idea to carry out such an integral (time-slice summation) at large distance is to model the correlation function by multi-hadron state {\it ansatz} or parametrizations of rho meson decay~\cite{Bernecker:2011gh,Chakraborty:2016mwy,DellaMorte:2017dyu}. The assumptions made may be more accurate when the pion mass is unphysically heavy or the size of the lattice box is small, but it is not clear how reliable these models are at physically light quark mass and large volume in which the rho meson becomes unstable and two-pion states become more dominant.

In addition, there are two different kinds of effects due to the finite four-dimensional volume. One effect is from the finite extent in the temporal direction, which causes the backward state propagation (BSP) due to the periodic boundary condition in time. Another, more complicated, effect is the finite spatial volume effects. To compare results in two different spatial volumes for large enough time extent, so that BSP effect is exponentially suppressed and becomes negligible, we extended the time extent, $N_t$ of a  gauge configuration, $U_\mu(x,t)$, by a factor of two by concatenating two identical lattices together in time direction,
\[
U^\text{ext}_\mu(x,t) =
\begin{cases}
  U_\mu(x,t) , ~~~ (0\leq t/a < N_t)\\
  U_\mu(x,t-N_t) , ~~~ (N_t\leq t/a < 2 N_t-1)
\end{cases}~.
\]
which is identical to the utilization of combining the quark propagators with periodic and anti-periodic boundary condition in temporal direction onto vector-vector current correlator. By comparing the t-integral on the original lattice, $U_\mu(x,t)$, and on the extended one, $U^\text{ext}_\mu(x,t)$, we will observe a significant effect of the BSP contribution to $a_\mu^{\rm hvp}$.

\section{Lattice set-up and its parameter}\label{sec:lattice_param}

In this paper, we use gauge configurations of a stout smeared non-perturbatively $\mathcal O(a)$ improved Wilson fermion in $N_f=2+1$ on Iwasaki gauge action with $\beta=1.82$ at the physical point (see Table~\ref{tab:lat_param}). PACS collaboration have generated it on two different volumes $L/a=64$ and 96, corresponding to 5.4 fm$^4$ and 8.1 fm$^3$, at a cut-off scale $a^{-1}=2.332(18)$~\cite{Ishikawa:2015fzw,Ishikawa:2015rho}.

In the measurement of vector-vector current correlator, we apply the AMA technique~\cite{Blum:2012uh,Blum:2012my,Shintani:2014vja} to boost the statistical accuracy. AMA is defined with the master formula for the measurement of target observable $O$, which is vector-vector current correlator in this case,
\begin{eqnarray}
  O^{\rm AMA} = \frac{1}{N_{\rm org}}\sum_{f\in G}^{N_{\rm org}}
    \Big[O^{{\rm (org)}\,f} - O^{{\rm (appx)}\,f}\Big] + \frac{1}{N_G}\sum_{g\in G}^{N_G} O^{{\rm (appx)}\,g},
\end{eqnarray}
with covariant transformation $g\in G$ under subset of its symmetry $G$. Here $G$ corresponds to translational symmetry and its size is $N_G$ for approximation and $N_{\rm org}$ for original. In \cite{vonHippel:2016wid}, one of the authors have developed the highly optimized AMA using Schwartz Alternative Procedure (SAP) deflation preconditioning~\cite{Luscher:2003qa,Luscher:2007se}. From a knowledge of tuning parameter to reduce the computational cost of approximation in AMA, SAP deflation makes an achievement of the high performance in the measurement on PACS configurations compared to low-mode deflation (see Appendix~\ref{appx:AMA}). Since for this study we need highly accurate lattice data of vector-vector current correlator in infrared region, we tune the parameter to generate the deflation field more efficiently in large time-slice. Using the limited number of gauge configurations to $<200$, separating 10 (20) and 40 HMC trajectory in $64^4$ ($64^3\times128$) and $96^4$ lattice ensembles respectively, we measure the approximation $O^{\rm (appx)\,g}$ with the $N_G\sim\mathcal O(10^3)$ as different source points, and its total statistics is thus $\mathcal O(10^6)$ we can achieve. Note that in error analysis we use the 5 (2) jackknife bin size for $64^4$ ($64^4\times128$) lattice ensemble, in which autocorrelation is small when using more than 40 HMC trajectory in PACS configurations~\cite{Ishikawa:2015rho}. From the practical point of view, aiming for $N_G\sim\mathcal O(10^3)$, we tune the parameter of approximation to be small $\Delta r$~\cite{Blum:2012uh,Blum:2012my,Shintani:2014vja}, in which $O^{\rm (appx)}$ and $O^{\rm (org)}$ are strongly correlated, as $\Delta r\lesssim N_G/2\sim\mathcal O(10^{-4})$ for the scaling of statistical error close to $1/\sqrt{N_G}$. 

In Table~\ref{tab:ama_param}, we show the detail of parameters to generate approximation $O^{\rm (appx)}$ in AMA on each gauge ensemble. In the computation of $O^{\rm AMA}$, we use a method with fixed number of iteration of General Conjugate Residual (GCR) solver with SAP deflation as used in \cite{vonHippel:2016wid}. In a generation of SAP deflation field, the domain block size, the number of SAP cycle $n_{\rm cy}$ and the number of deflation vector are tuned as in Table~\ref{tab:ama_param}. SAP is used in not only preconditioning of GCR, but also generation of deflation field overlapping with low-mode dominance by smoothing technique (inexact deflation~\cite{Luscher:2007se}). 

\begin{table}
  \begin{center}
    \caption{Table of parameters of PACS gauge ensembles. $L$ denotes spatial length and $T$ denotes temporal length. In Wilson-clover fermion, $K_l$ and $K_s$ denotes kappa value for light (up and down) quark flavor and strange quark flavor respectively. $(^*)$This is a kappa for valence quark to shift the pion mass to be $0.135$ GeV. For sea quark, we use the reweighting method to be on the unitary point.}\label{tab:lat_param}
    \begin{tabular}{ccccccccc}
      \hline\hline
      $L/a$ & $T/a$ & $K_{l}$ & $K_s$ & $m_\pi$(GeV) & $m_K$(GeV) & $m_\pi L$ & configs. \\ 
      \hline
      96[8.1 fm] & 96[8.1 fm] & 0.126117 & 0.124790 & 0.1461(4) & 0.5242(3) & 6.0 & 50 \\
      \hline
      64[5.4 fm] & 64[5.4 fm] & 0.126117 & 0.124902 & 0.1385(9) & 0.5004(4) & 3.9 & 187 \\
      & & 0.126119$^*$ & 0.124902 & 0.1354(9) & 0.4999(4) & 3.7 & 87 \\
      \hline\hline
    \end{tabular}
  \end{center}
\end{table}

\begin{table}
  \begin{center}
    \caption{Table of parameters in our analysis using AMA on each ensemble. ``Domain block'' is a size of domain in SAP. $n_{\rm cy}$ is a number of cycle of SAP. $N_s$ denotes the number of deflation vector. ``Stop iter.'' column is a fixed number of iteration of GCR with SAP deflation. $r_{\rm src}^{\rm min}$ denotes the minimum separation of source point between $O^{{\rm (appx)}\,g}$. ``Meas.'' column is a total number of measurement for $O^{\rm (AMA)}$, which consists with the number of configuration times $N_G$ measurements of $O^{\rm (appx)}$. The last two rows are in the ensemble whose temporal length is extended into double size by duplication.}\label{tab:ama_param}
    \begin{tabular}{ccccccccc}
      \hline\hline
      \multirow{2}{*}{$L/a$} & \multirow{2}{*}{$T/a$} & \multirow{2}{*}{quark} & Domain & \multirow{2}{*}{$n_{\rm cy}$} & \multirow{2}{*}{$N_s$} & \multirow{2}{*}{Stop iter.} & \multirow{2}{*}{$r_{\rm src}^{\rm min}$ fm} & \multirow{2}{*}{Meas.} \\
      & & & block \\
      \hline
      \multirow{2}{*}{96[8.1 fm]} & \multirow{2}{*}{96[8.1 fm]} & light & \multirow{2}{*}{6$^4$} & \multirow{2}{*}{5} & 40 & \multirow{2}{*}{5} & 1.43 &  102,141\\
       &  & strange & & & 30 & & 2.03 & 3,382 \\
      \hline
      \multirow{4}{*}{64[5.4 fm]} & \multirow{2}{*}{64[5.4 fm]} & light & \multirow{4}{*}{4$^4$} & \multirow{4}{*}{5} & \multirow{4}{*}{30} & \multirow{4}{*}{5} & 0.68 & 409,966\\
      & & strange & & & & & 1.35 & 6,247\\
      & \multirow{2}{*}{128[10.8 fm]} & light & & & & & 0.68 & 157,250\\
      & & strange & & & & & 2.71 & 1,376\\
      \hline
      \multirow{4}{*}{64[5.4 fm]} & \multirow{2}{*}{64[5.4 fm]} & light & \multirow{4}{*}{4$^4$} & \multirow{4}{*}{5} & \multirow{4}{*}{30} & \multirow{4}{*}{5} & \multirow{4}{*}{0.68} & \multirow{2}{*}{425,991}\\
      & & (reweight) \\
      & \multirow{2}{*}{128[10.8 fm]} & light & & & & & & \multirow{2}{*}{111,360}\\
      & & (reweight) \\
      \hline\hline
    \end{tabular}
  \end{center}
\end{table}

\section{Numerical results}\label{sec:result}
\subsection{Analysis of vector-vector current correlator}
\begin{figure}
  \begin{center}
    \includegraphics[width=75mm]{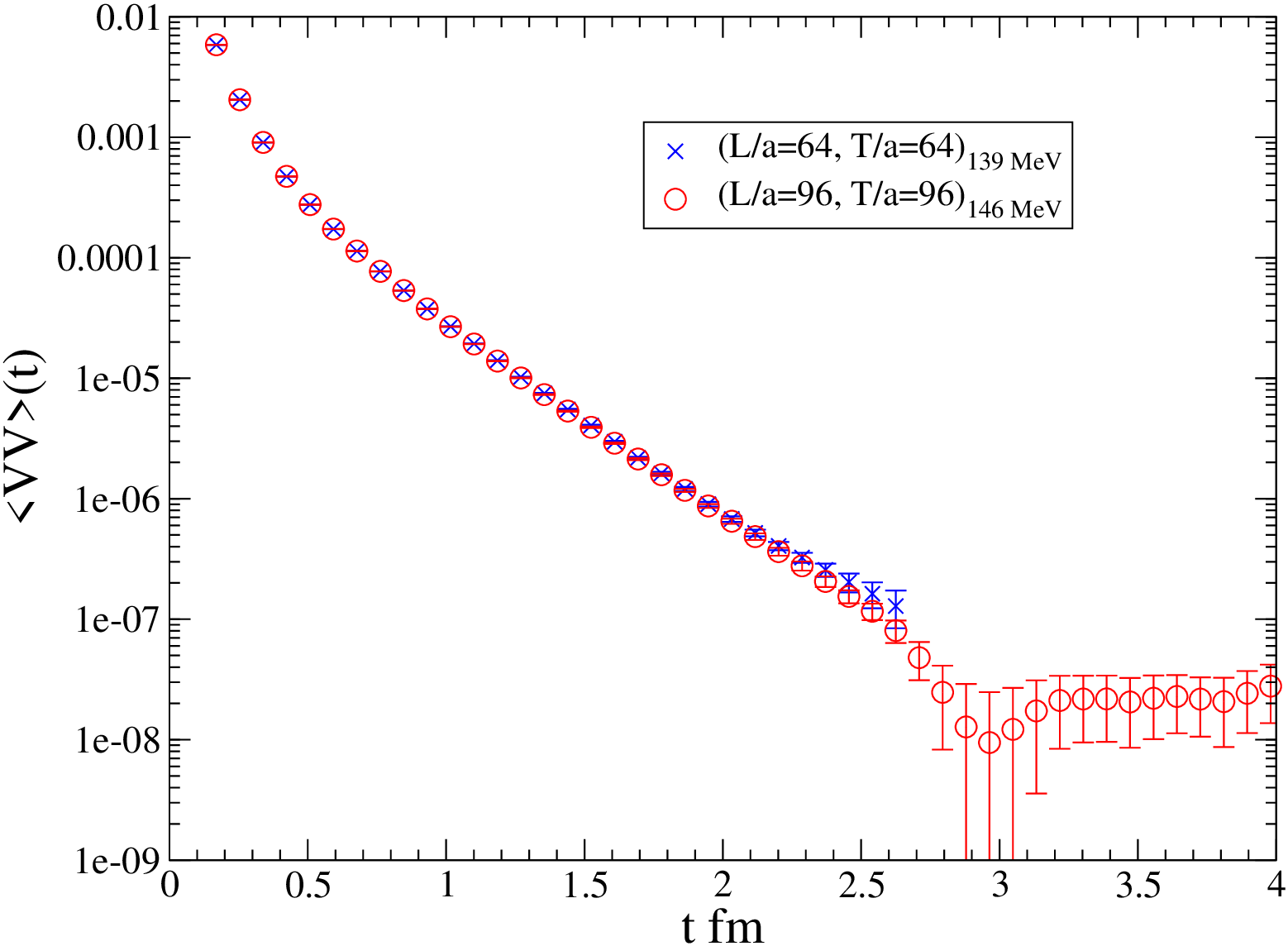}
    \hspace{4mm}
    \includegraphics[width=70mm]{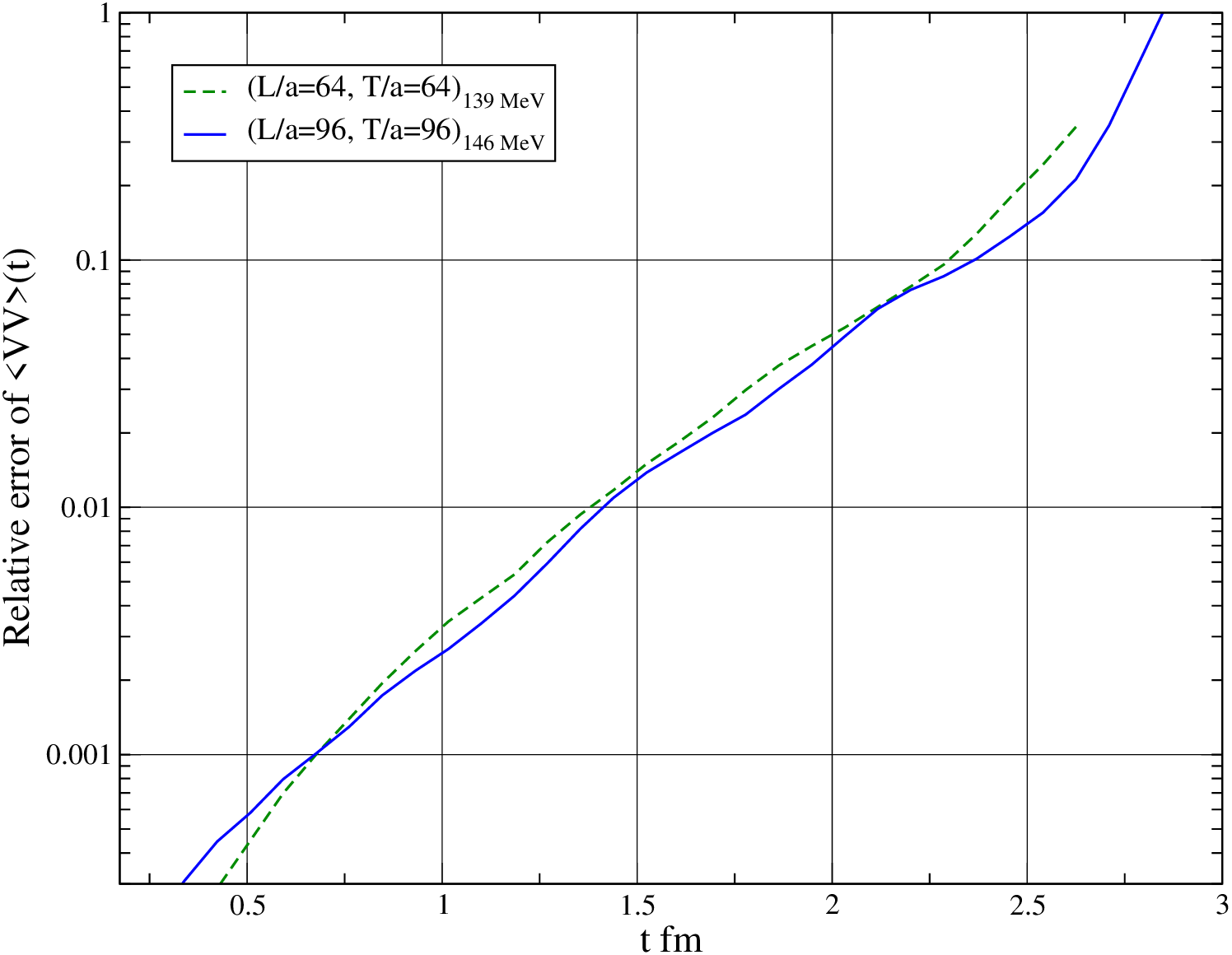}
    \caption{(Left) The vector current two point correlation function as a function of time-separation in fm unit. (Right) Relative error of vector-vector correlator as a function of time-slice in fm unit. Different lines denote results of the different gauge ensembles.}\label{fig:vvt_corr}
  \end{center}
\end{figure}

First, we show the time-separation dependence of vector current two point correlation function, $C(t)$, from short to long distance in Figure~\ref{fig:vvt_corr}. For the computation of $a_\mu^{\rm hvp}$ using eq.~(\ref{eq:amu_TMR_lat}), we need to know the precise value of $C(t)$ in large $t$ region. Our high-statistics result boosted by AMA method show a statistically significant signal beyond $t=2.7$ fm, which is the longest temporal separation for our smaller $64^4$ lattice, and it thus allows us to compare large $t$ behavior of $C(t)$ on  different volumes.

\begin{figure}
  \begin{center}
    \includegraphics[width=100mm]{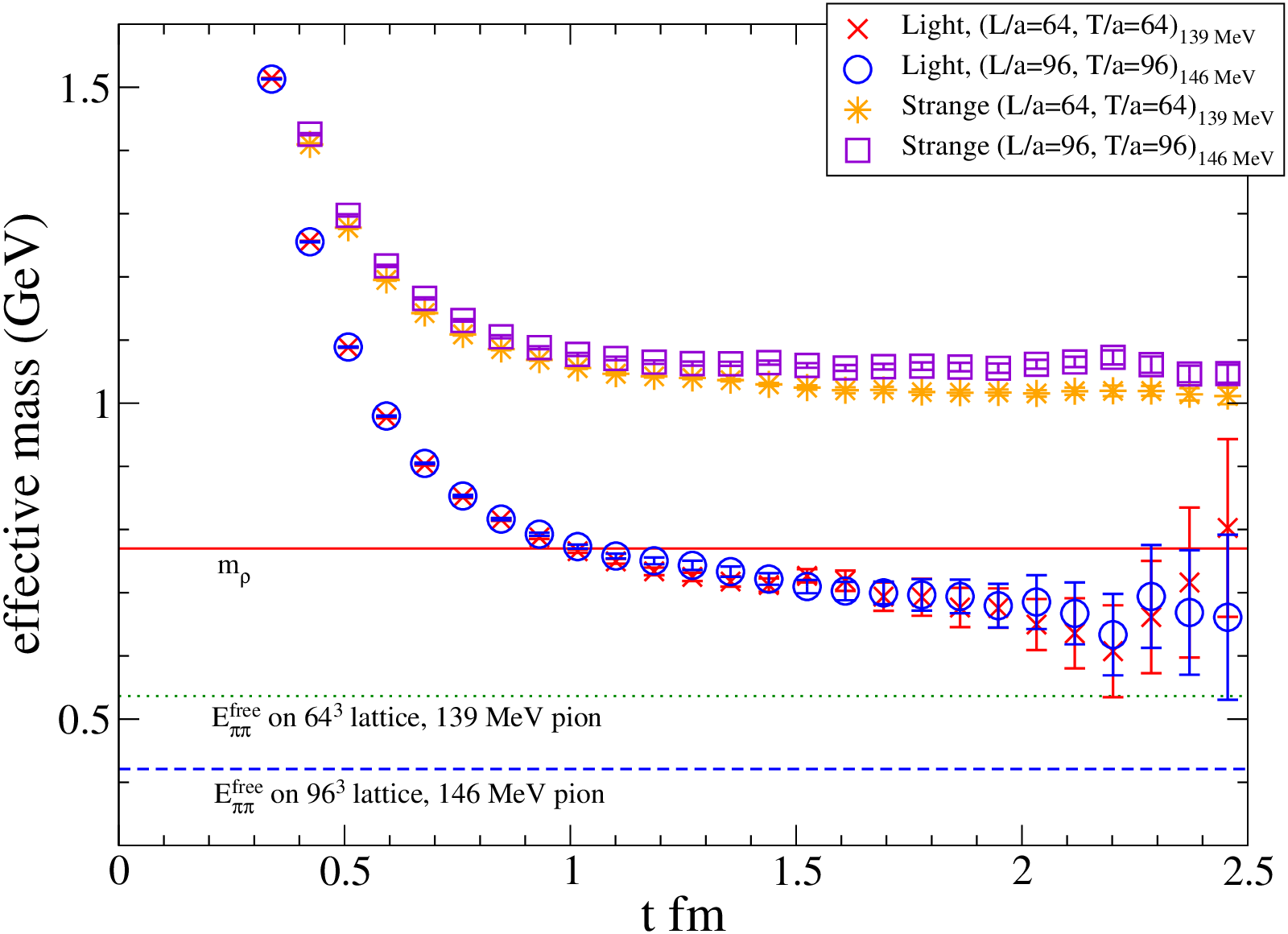}
    \caption{Effective mass plot of vector-vector current correlator at light and strange quark flavor. Different symbols denote the results in different flavors and lattice sizes. The straight solid-line shows the experimental value of  rho meson mass, and dashed-line and dotted-line show the 2 free pion energy in $m_\pi=0.146$ GeV on $96^4$ lattice and $m_\pi=0.139$ GeV on $64^4$ lattice, respectively.}\label{fig:effm_vv}
  \end{center}
\end{figure}

It is also noteworthy that energy of (non-interacting) two light pions ($m_\pi\simeq$ 0.15 and 0.14 GeV)  in our both gauge ensembles with large volumes (L=8.1 and 5.4 fm) is smaller than rho meson threshold : $m_\rho > E_{\pi\pi}>2\sqrt{m_\pi^2+(2\pi/L)^2}\approx0.42$ and 0.54 GeV. In fact, the effective energy of $C(t)$ in Figure~\ref{fig:effm_vv} are clearly smaller than rho meson resonance energy, $\simeq 770$ MeV,  at $t>1.2$ fm.

\begin{figure}
  \begin{center}
    \includegraphics[width=100mm]{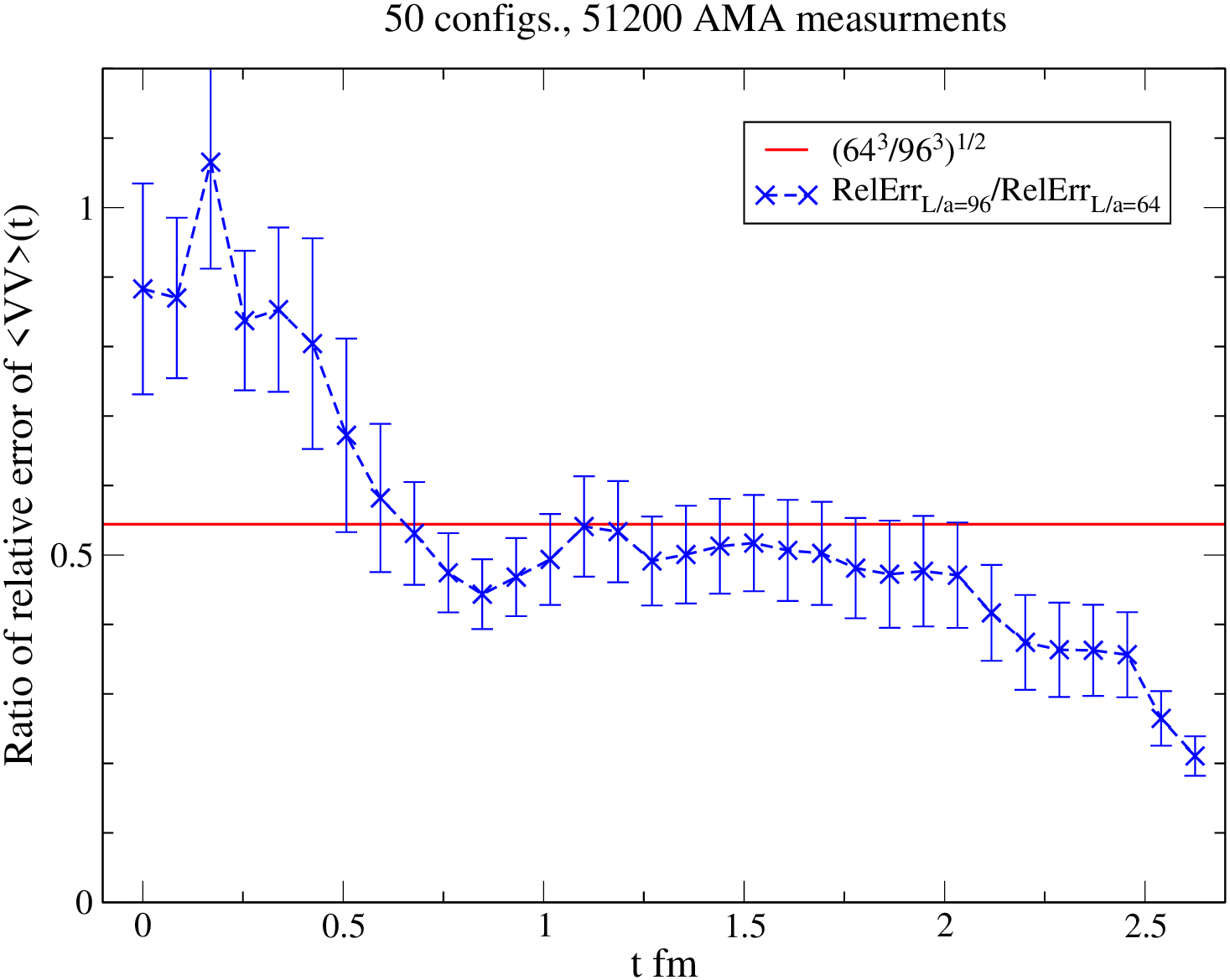}
    \caption{The ratio of relative error of vector-vector current correlator in $L/a=96$ and $L/a=64$ on the same number of measurements.}\label{fig:relerr_ratio}
  \end{center}
\end{figure}

Right panel of Figure~\ref{fig:vvt_corr} is a relative statistical error of correlation functions for each volume, which shows that the $L/a=96$ errors are comparable with $L/a=64$, even though the number of measurements in $L/a=96$ is about 4 times smaller than $L/a=64$. To see more details of such an error reduction for the larger volume, we plot the ratio of the two relative errors from $L/a=96$ and $L/a=64$ for the same number of measurements, 51,200 on 50 gauge configurations in Figure~\ref{fig:relerr_ratio}. In this plot, error-bar is obtained from error-of-error analysis
in which we compute the standard error of error from 10 sampling of standard deviations within 50 ensembles, {\it i.e.} splitting 10 of ensemble-errors obtained with 5 ensembles from 50 gauge ensembles. One can see that the ratio at $t$ between  0.5 fm and 2 fm is close to the square-root of spatial volume ratio, $\sqrt{64^3/96^3} \simeq 0.544$. Beyond 2 fm, the relative error further decreases due to large statistical error of smaller lattice (64$^4$) in which the BSP becomes significant as we will discuss in the next subsection. This statistical advantage on large lattice volume is an encouraging observation for HVP calculation
\footnote{While we don't have definite theoretical explanation as to this error reduction in large volume, the lighter pion mass for smaller box ($64^4$) would naturally cause this reduction. However this may be unlikely the explanation of the constant behavior seen at $t\in$ [1 fm, 2 fm] in Figure~\ref{fig:relerr_ratio}. As studied in \cite{DelDebbio:2005qa}, the stability of spectral gap in Wilson type fermion in large volume may be other possibility. Further detail of such an error reduction using more larger lattice size at exactly same pion mass is much interesting.}.

\subsection{Computation of $a_\mu^{\rm hvp}$}
\begin{figure}
  \begin{center}
    \includegraphics[width=75mm]{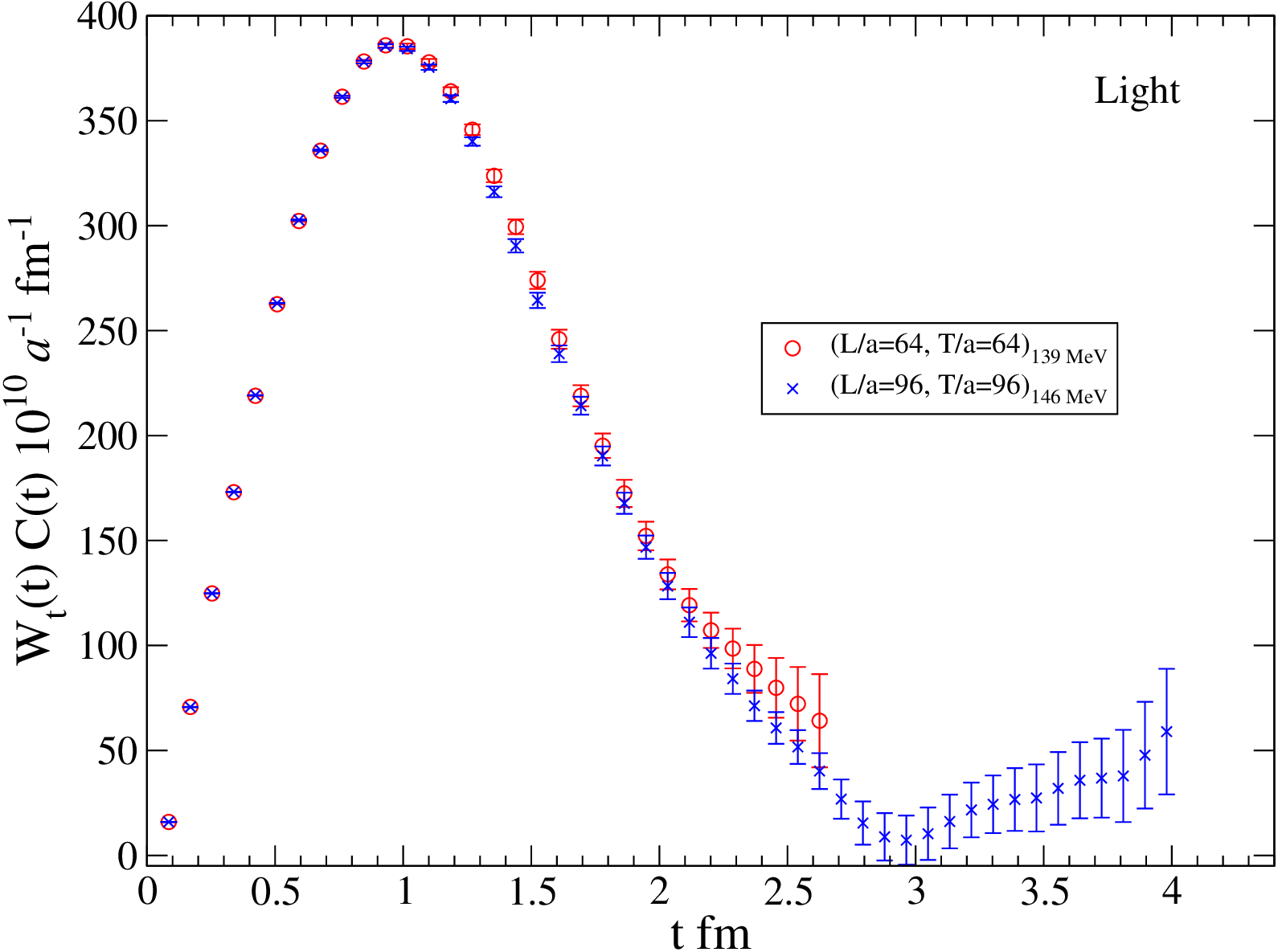}
    \hspace{5mm}
    \includegraphics[width=75mm]{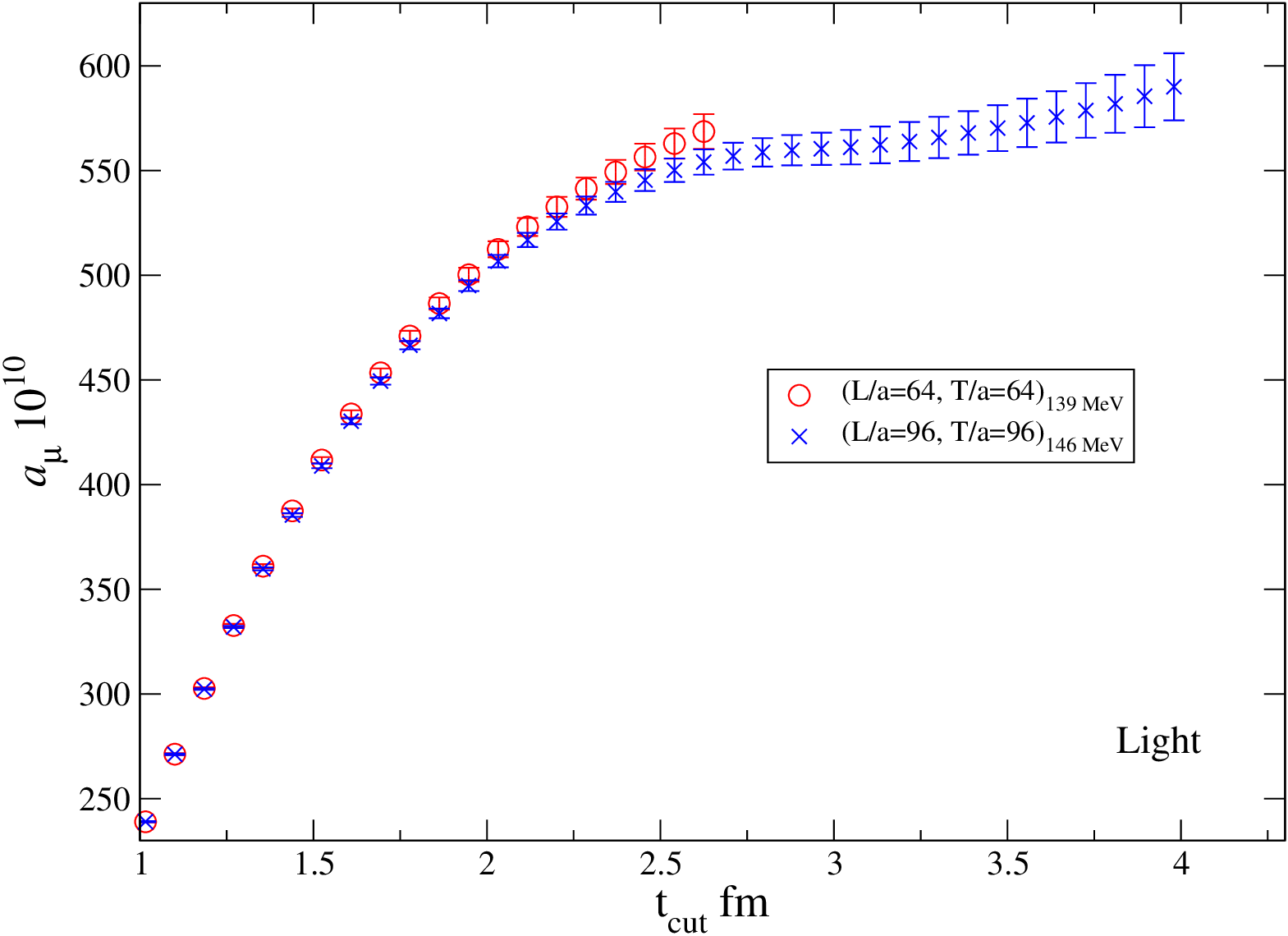}
    \caption{(Left) Integrand of $[a_\mu^{\rm hvp}]_{\rm lat}$  in Eq.~(\ref{eq:amu_TMR_lat}) divided by lattice spacing as a function of time-slice in physical unit. Different symbols denote the results in each gauge ensemble at light flavor. (Right) Time-slice summation for $[a_\mu^{\rm hvp}]_{\rm lat}$ up to $t_{\rm cut}$ at light quark flavor. Different symbols denote the data of each gauge ensemble. }\label{fig:amu_tdep}
  \end{center}
\end{figure}

In this section we present the volume dependence for the integrand of $[a_\mu^{\rm hvp}]_{\rm lat}$ (see Eq.~(\ref{eq:amu_TMR_lat})) and its time-slice summation. The integrand at time-slice $t$ and time-slice summation up to $t_{\rm cut}$ for the light quark contribution on each lattice volume is compared in Figure~\ref{fig:amu_tdep}. The light quark contribution $[a_\mu^{\rm hvp}]^l_{\rm lat}$ is dominated in $[a_\mu^{\rm hvp}]_{\rm lat}$ (strange quark contribution to $[a_\mu^{\rm hvp}]_{\rm lat}$ is a few percent magnitude due to its larger mass and $1/5 = (e_s/e_l)^2$ factor of electric charge ratio, see section~\ref{sec:strange}). Compared between $L/a=96$ and $L/a=64$ lattice, the shape of its integrand is similar to each other until $t=1$ fm, and it then appears that $64^4$ lattice data is slightly larger than $96^4$ lattice data at $t\simeq 1.3$ fm. In the right panel of Figure~\ref{fig:amu_tdep}, however, we observe its time-slice summation is not significantly different even at $t_{\rm cut}=2.5$ fm. It means there is not so large FV effect on $L/a=64$, which is similar order of magnitude to its statistical fluctuation. To robustly estimate the magnitude of FV effect between $L/a=64$ and $L/a=96$, we take into account the correction of its mass difference, 146 MeV ($L/a=96$) and 139 MeV pion ($L/a=64$) as shown in section~\ref{sec:mass_correction}. 

\begin{figure}
  \begin{center}
    \includegraphics[width=75mm]{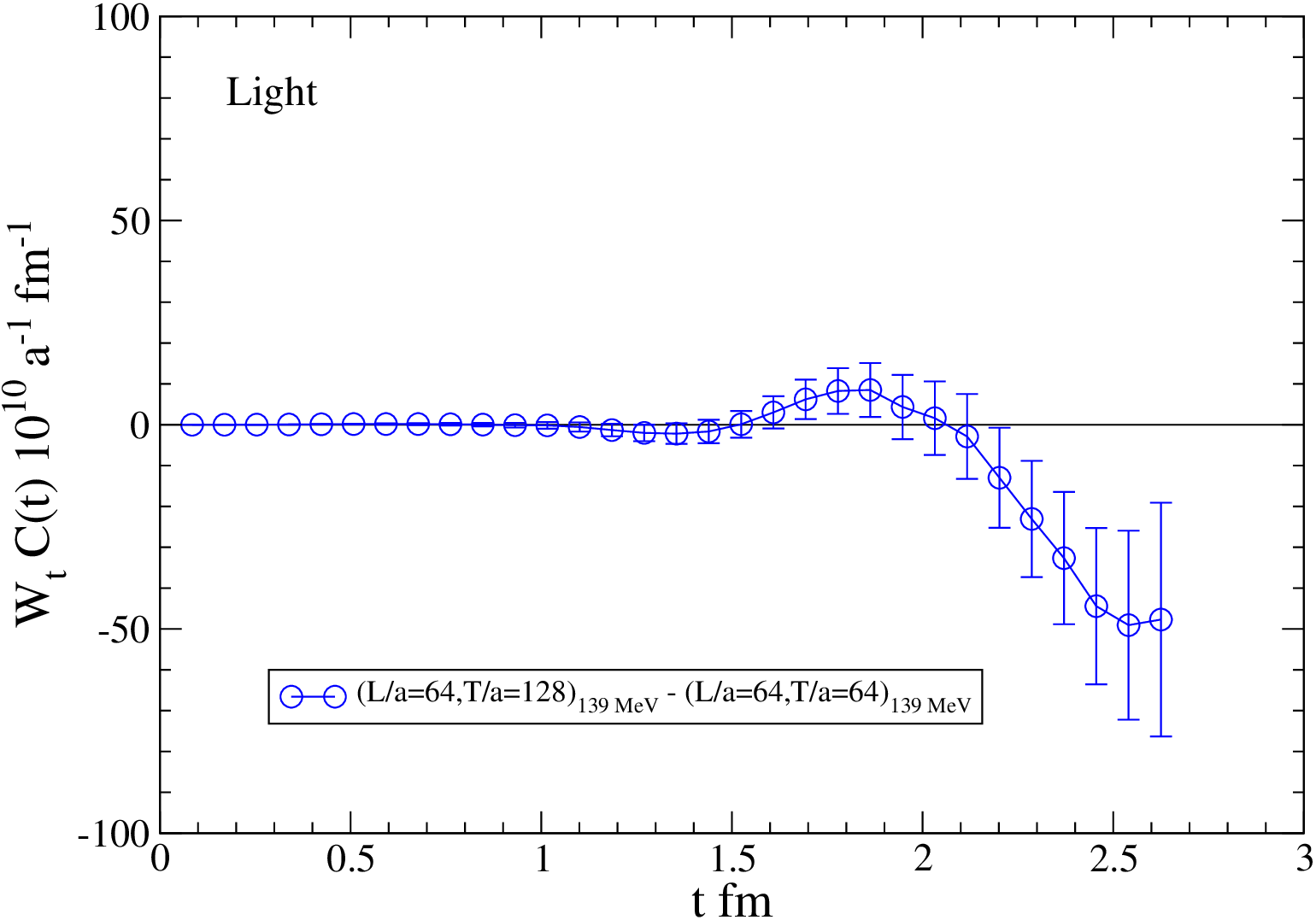}
    \hspace{5mm}
    \includegraphics[width=72mm]{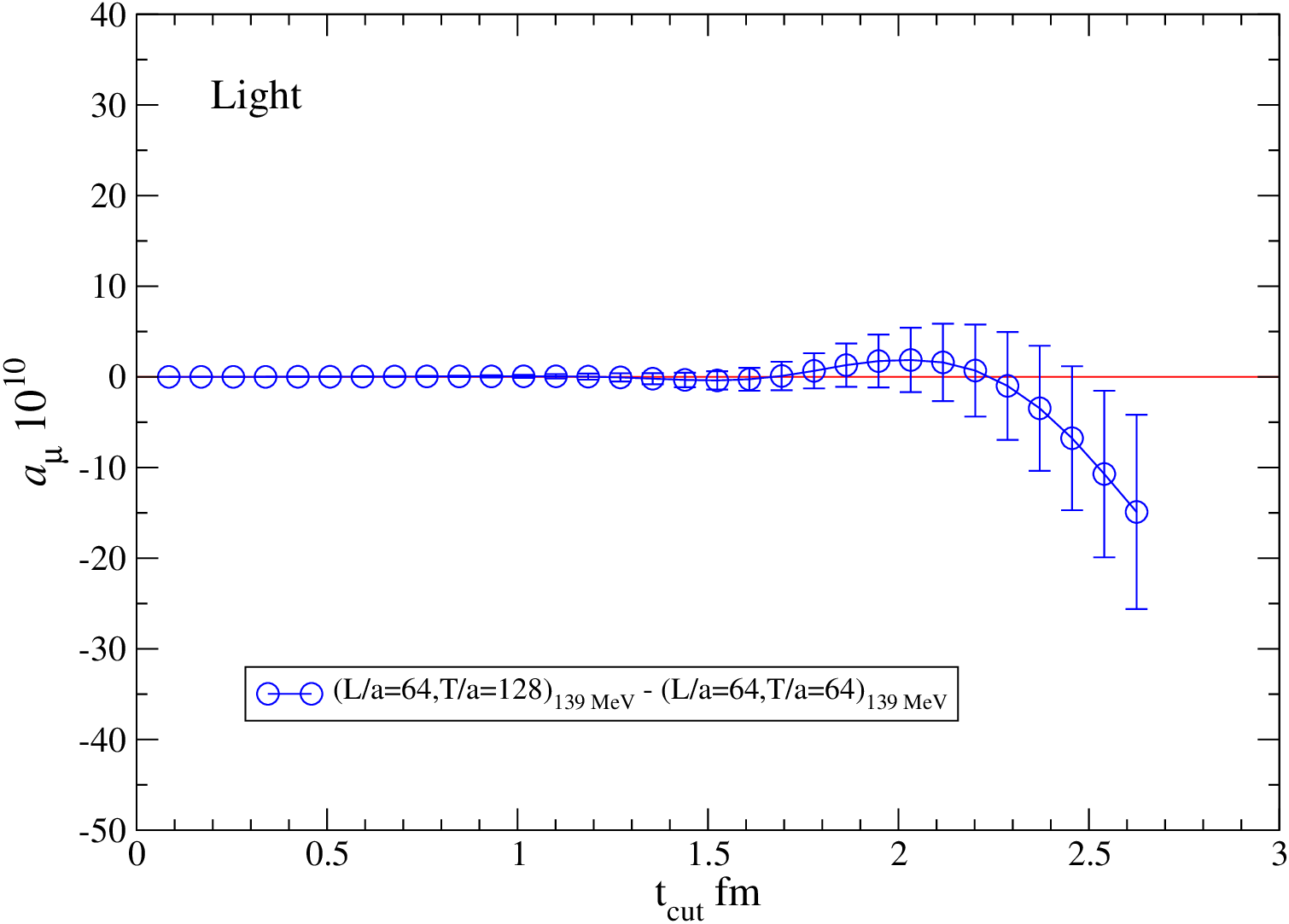}
    \caption{Difference of integrand divided by lattice spacing (left) and time-slice summation up to $t_{\rm cut}$ (right) from $L/a=64,T/a=128$ lattice to $L/a=64,T/a=64$ lattice at light quark flavor.}
    \label{fig:diff_64_128}
  \end{center}
\end{figure}

In order to observe the appearance of BSP into the integrand and its time-slice summation, we compare two sizes of temporal extension on $L/a=64$ lattice, one is the original size as $T/a=64$ and another is the extended one as $T/a=128$, in Figure~\ref{fig:diff_64_128}. From $t\approx2.4$ fm, the BSP contribution to the integrand significantly appears, and time-slice summation is then maximally affected by BSP about 4\% contribution at $t_{\rm cut}=2.6$ fm. To avoid the unwanted contribution to BSP on $T/a=64$ lattice, $t_{\rm cut}=2$ fm is safe. 

\begin{figure}
  \begin{center}
    \includegraphics[width=70mm]{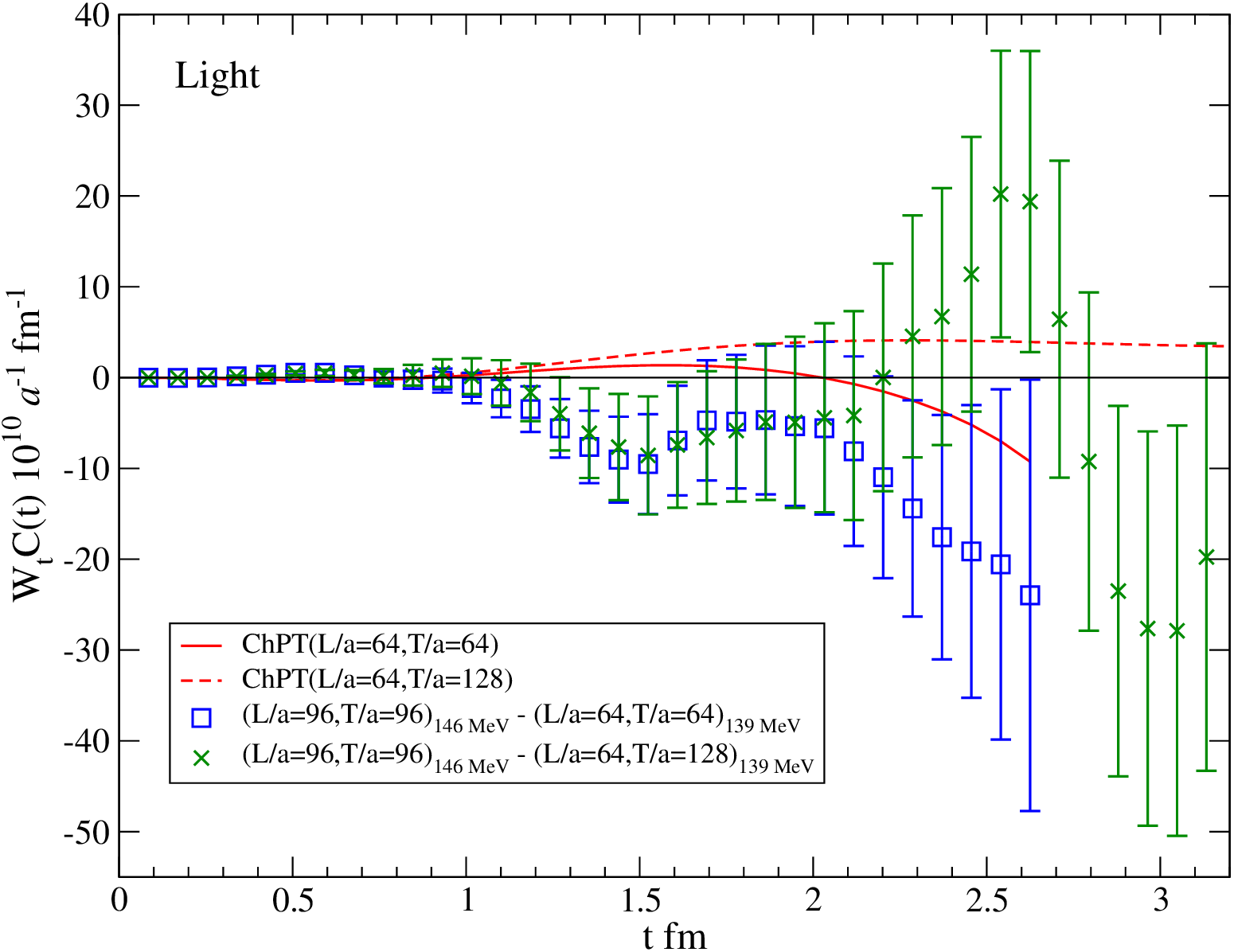}
    \hspace{5mm}
    \includegraphics[width=73mm]{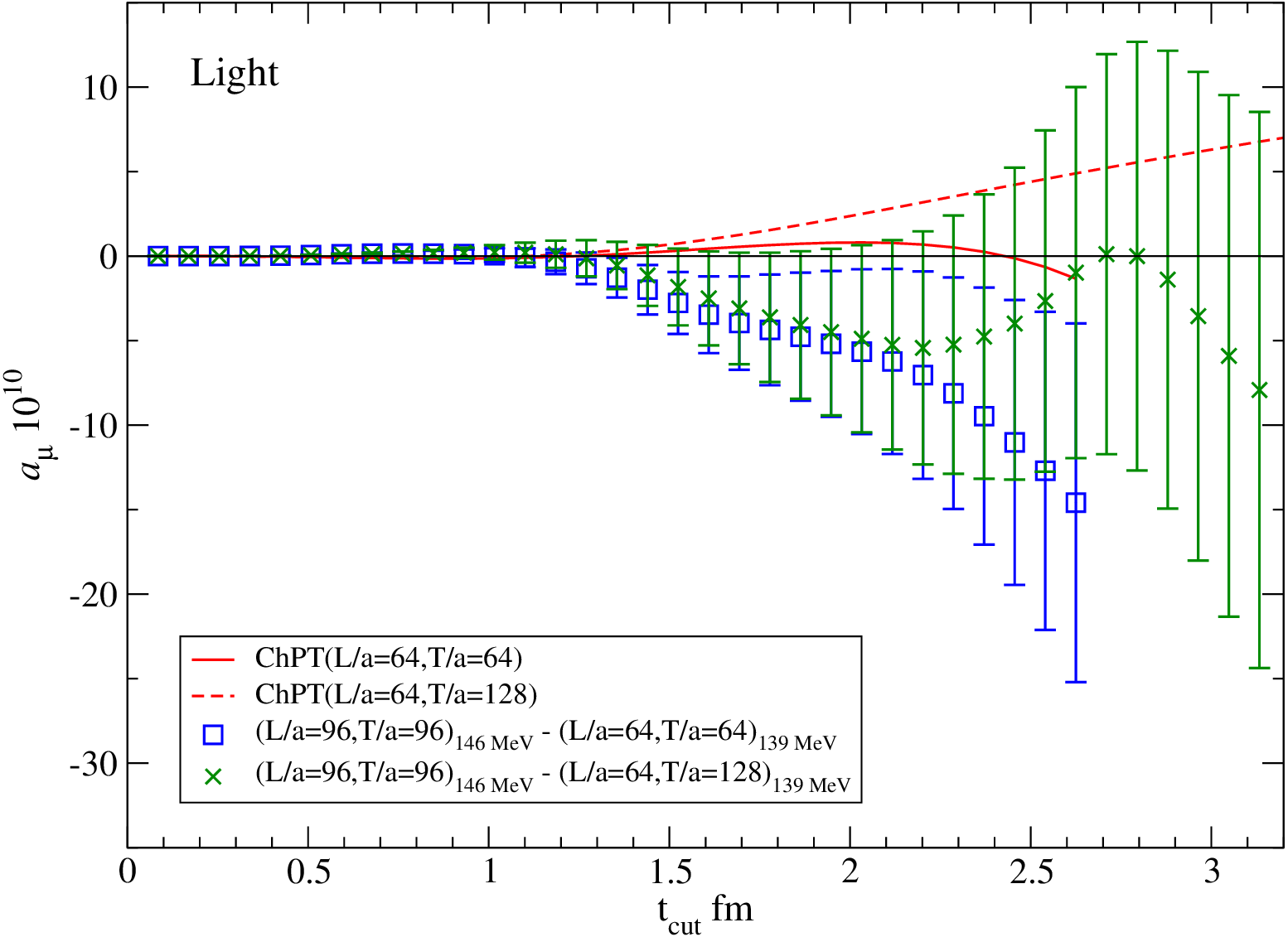}
    \caption{(Left) Plot of difference from integrand on $96^4$ lattice with 146 MeV pion to integrand on $64^4$ lattice (squared symbol), and $64^3\times128$ lattice (cross symbol), with 139 MeV pion at light quark flavor. (Right) The difference of $[a_\mu^{\rm hvp}]^l_{\rm lat}$ up to $t_{\rm cut}$ on different lattice size using same symbol with left panel at light quark flavor. Note that there has not been mass correction between $96^4$ and $64^4$ lattice yet. See in section \ref{sec:mass_correction}. Solid (dashed) lines in both figures denote the leading order of ChPT on $64^4$ ($64^3\times128$) lattice.}\label{fig:amu_diff}
  \end{center}
\end{figure}

As in Figure~\ref{fig:amu_diff}, showing the difference between $L/a=96$ and $L/a=64$ lattice at light quark flavor, the integrand is excellently consistent until $t=1$ fm. From $t=1$ fm to 2 fm, slightly negative discrepancy appears, while it is less than 10$\times 10^{-10}$ for $[a_\mu^{\rm hvp}]_{\rm lat}^l$ at $t_{\rm cut}\approx2$ fm. At $t>2$ fm on $T/a=64$, since there is a significant appearance of BSP as positive effect, which has been observed in Figure~\ref{fig:diff_64_128}, the discrepancy between data of $T/a=64$ and $T/a=128$ also appears. Note that, as mentioned before, discrepancy between data of $96^4$ and $64^4$ lattice ensembles may be due to not FV effect but a mass correction to slight pion mass difference $\sim$7 MeV between two ensembles. Compared to the leading order of chiral perturbation theory (ChPT)\cite{Aubin:2015rzx,Golterman:2017njs}, which indicates that integrand increases from heavy to light mass ($m_\pi=146$ MeV$\rightarrow$ 139 MeV), while decreases from large to small volume ($L/a=96\rightarrow 64$), such a discrepancy becomes small by the cancellation of both effects. In Figure~\ref{fig:amu_diff} and after, we present the comparison with ChPT on the corresponding box sizes. One can see that the discrepancy of $[a_\mu^{\rm hvp}]^l_{\rm lat}$ between 96$^4$ and 64$^4$ lattice ensembles slightly differs from ChPT lines including those signs, besides it is mostly overlapping with 1-$\sigma$ statistical error bar. Later further comparison will discuss after analysis of mass difference on the same volume and volume difference on the same mass with extrapolation using mass-reweighted lattice data. Note that ChPT line at $t>2$ fm with $T/a=64$ has negative curvature due to BSP effect, which is consistent behavior with lattice data. 

\subsubsection{Upper and lower bound}

Here we estimate the bound of $[a_\mu^{\rm hvp}]^l_{\rm lat}$ should be satisfied at $t_{\rm cut}$ as argued in \cite{LehnerTalkLGT16,Borsanyi:2016lpl,Borsanyi:2017zdw,Blum:2018mom}. The upper-bound under the assumption of free two-pion state dominance is given as,
\begin{eqnarray}
  \big[a_\mu^{\rm hvp}\big]_{\rm upper} &=& \big[a_\mu^{\rm hvp}\big]_{\rm lat}(t_{\rm cut}) + \sum_{t/a=t_{\rm cut}/a}^\infty C(t_{\rm cut})\frac{e^{-E_{\pi\pi}t}}{e^{-E_{\pi\pi}t_{\rm cut}}+e^{-E_{\pi\pi}(T-t_{\rm cut})}}W_t(t),
  \label{eq:amu_upper}
\end{eqnarray}
where $E_{\pi\pi}$ represents the energy of free two-pion state, $E_{\pi\pi}=2\sqrt{m_\pi^2+(2\pi/L)^2}$. In fact, one can see from Figure~\ref{fig:effm_vv} the effective mass of vector-vector current correlator is still above $E_{\pi\pi}$ even at $t>1.5$ fm, and so that using the integrand switched to the single exponential function with $E_{\pi\pi}$ after $t=t_{\rm cut}$, Eq.~(\ref{eq:amu_upper}) is a restricted upper-bound for $a_\mu^{\rm hvp}$ in time-slice summation. On the other hand, the lower bound we take is two forms, 
\begin{eqnarray}
  \big[a_\mu^{\rm hvp}\big]_{\rm lower(0)} &=& \big[a_\mu^{\rm hvp}\big]_{\rm lat}(t_{\rm cut}),
  \label{eq:amu_lower}\\
  \big[a_\mu^{\rm hvp}\big]_{\rm lower(m_\rho)} &=& \big[a_\mu^{\rm hvp}\big]_{\rm lat}(t_{\rm cut}) + \sum_{t/a=t_{\rm cut}/a}^\infty C(t_{\rm cut})\frac{e^{-m_\rho t}}{e^{-m_\rho t_{\rm cut}}+e^{-m_\rho(T-t_{\rm cut})}}W_t(t).
  \label{eq:amu_lower2}
\end{eqnarray}
The first lower-bound in Eq.~(\ref{eq:amu_lower}) is consistent with $[a_\mu^{\rm hvp}]_{\rm lat}$ since we know the remnant integral from $t_{\rm cut}$ to infinity is positive contribution. Otherwise, the second lower-bound in Eq.~(\ref{eq:amu_lower2}), consists of the exponential function with the rho mass (0.775 GeV) from $t>t_{\rm cut}$, which is same form as the upper-bound of Eq.~(\ref{eq:amu_upper}) instead of $E_{\pi\pi}$. The second one is a more restricted bound, since an additional contribution of the rho state is taken into account. Figure~\ref{fig:effm_vv} which presents the lower exponent of the vector-vector current correlator than the rho mass at $t>1$ fm actually shows that such a restricted lower bound is reasonable for our data. 

\begin{figure}
  \begin{center}
    \includegraphics[width=74mm]{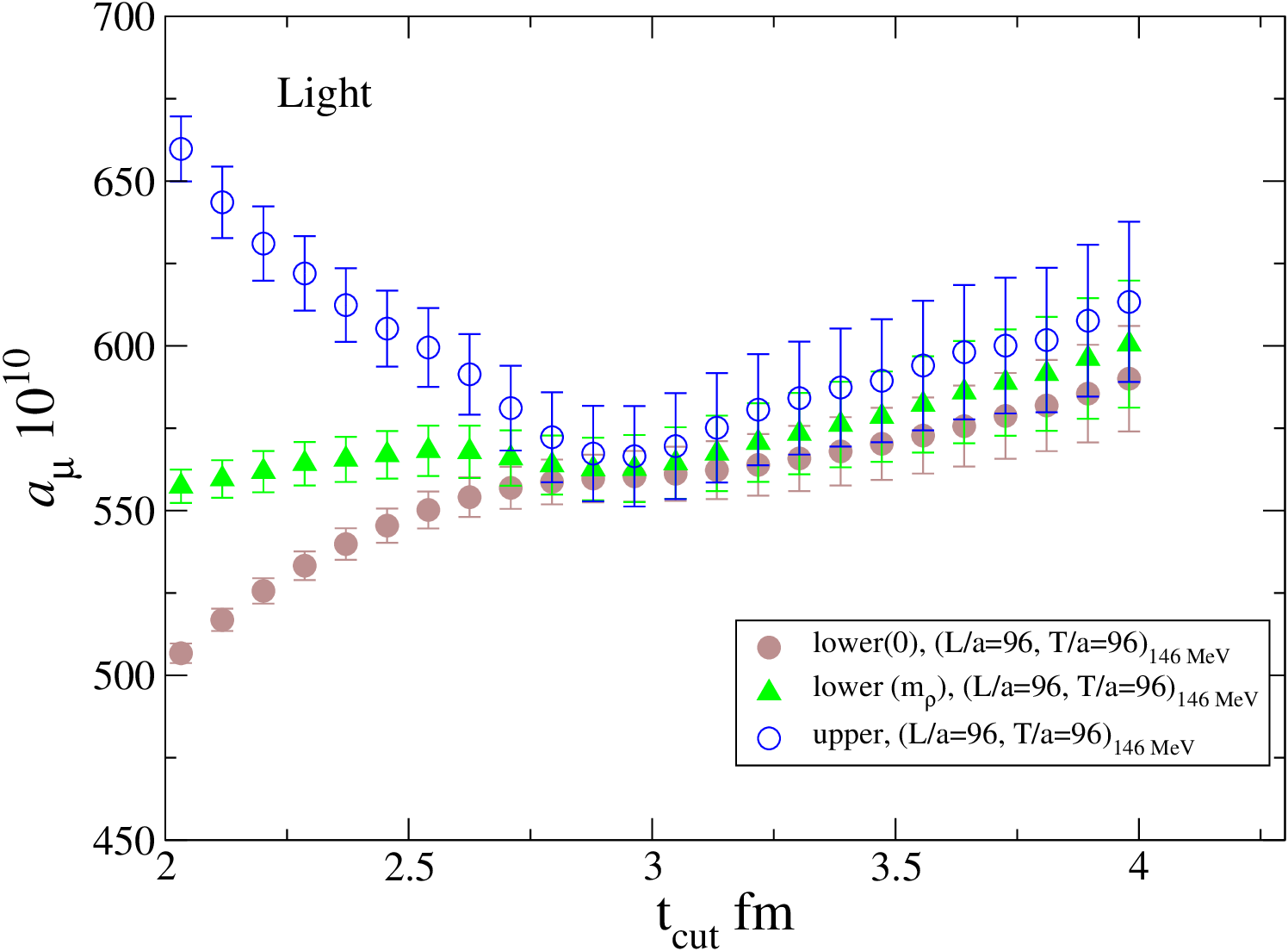}
    \hspace{5mm}
    \includegraphics[width=74mm]{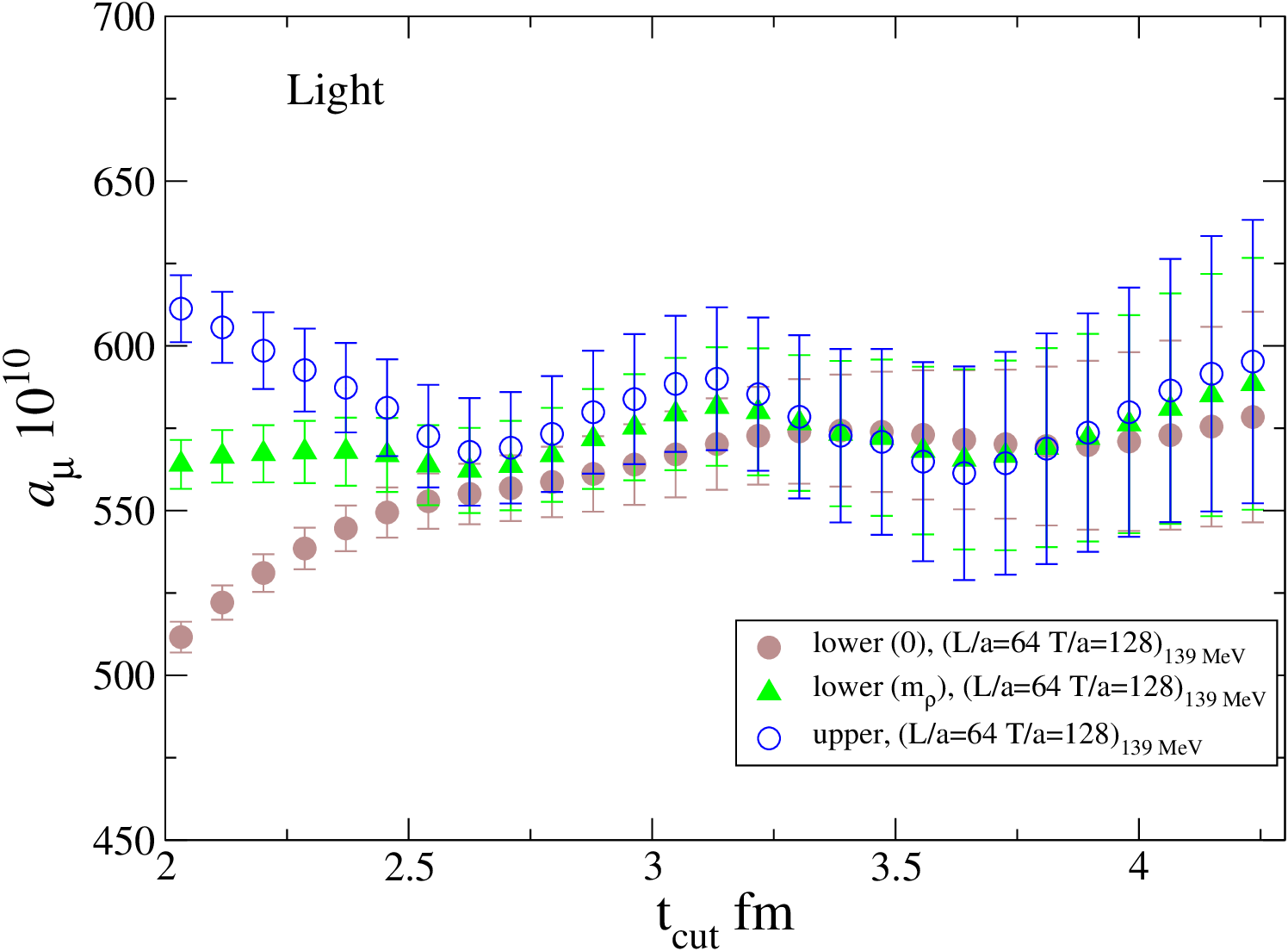}
    \caption{This plot shows the upper and lower bounds of $[a^{\rm hvp}_\mu]^l_{\rm lat}$ at each $t_{\rm cut}$ in light quark flavor on $96^4$ lattice (left) and $64^3\times128$ lattice (right). Filled circle- and triangle-symbol denotes the lower-bound defined in Eq.~(\ref{eq:amu_lower}) and Eq.~(\ref{eq:amu_lower2}) respectively. Open cross-symbol denotes the upper-bound in Eq.~(\ref{eq:amu_upper}). }\label{fig:amu_integ_bnd}
  \end{center}
\end{figure}

Figure \ref{fig:amu_integ_bnd} shows such a lower- and upper-bound on two lattice volumes, $L/a=96$ and $L/a=64$ lattices. At $t_{\rm cut}\approx 3$ fm, two bounds become consistent within 1 $\sigma$ statistical error. Since the statistical error of upper-bound is larger than two lower-bounds due to rather large fluctuation of $C(t_{\rm cut}$), the upper-bound at $t_{\rm cut}\approx3.0$ fm is regarded as a possible range of $[a_\mu^{\rm hvp}]^l$. In our analysis, we have 
\begin{eqnarray}
  \begin{array}{ccc}
   554 < \big[a_\mu^{\rm hvp}\big]^l_{\rm lat} \times 10^{10}< 586, & [\textrm{$L/a$=96 lattice with 146 MeV pion}],\\
   562 < \big[a_\mu^{\rm hvp}\big]^l_{\rm lat} \times 10^{10}< 609, & [\textrm{$L/a$=64 lattice with 139 MeV pion}],\\
  \end{array}
  \label{eq:amu_tsum_bnd}
\end{eqnarray}
in which the upper value of $\big[a_\mu^{\rm hvp}\big]^l_{\rm upper}$ and lower value of $\big[a_\mu^{\rm hvp}\big]^l_{\rm lower(m_\rho)}$ is consistent within 1-$\sigma$ statistical error. One can see two regions are mostly overlapping. 

\subsection{Strange quark contribution}\label{sec:strange}
\begin{figure}
  \begin{center}
    \includegraphics[width=100mm]{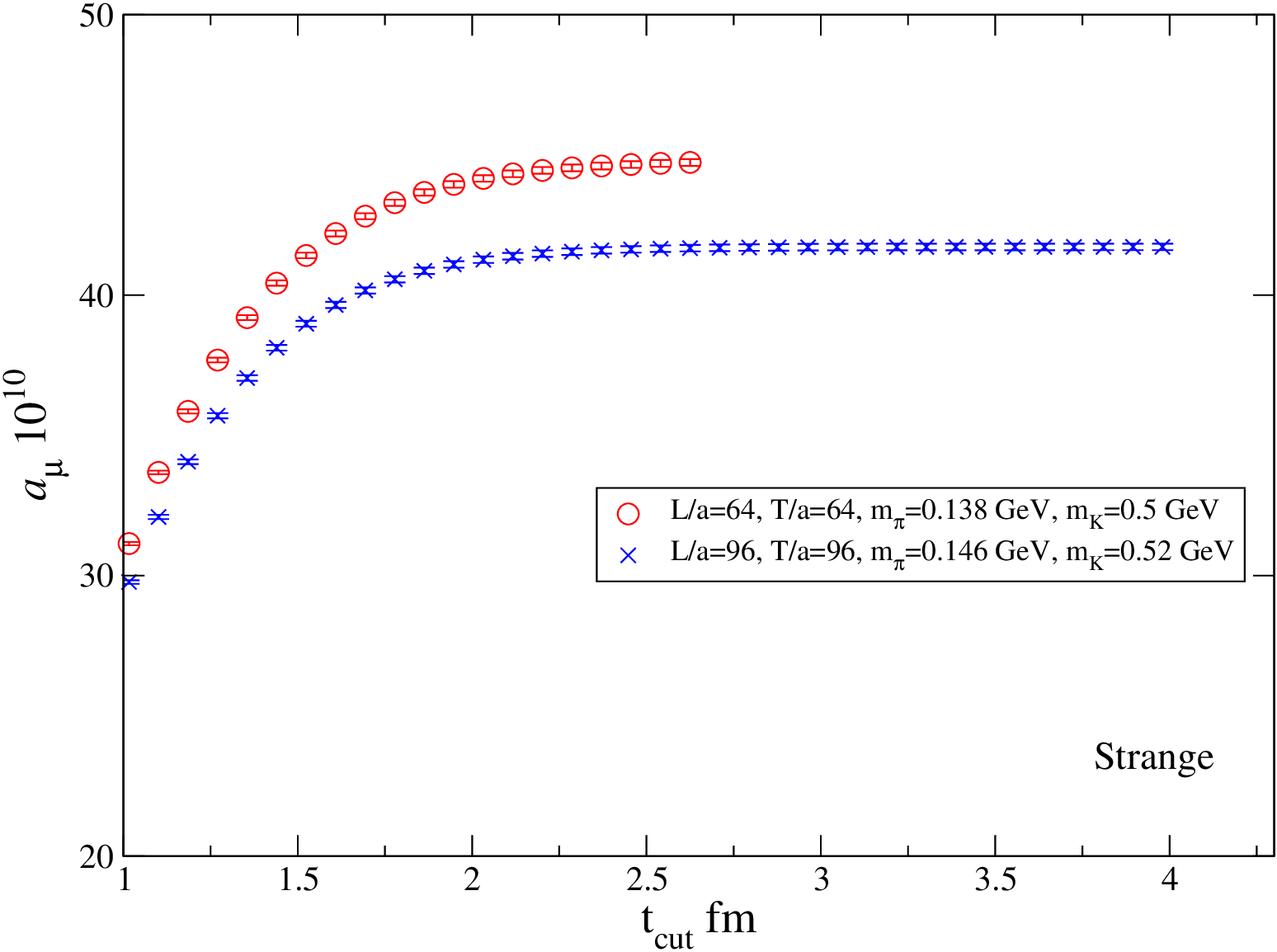}
    \caption{The time-slice summation of $[a_\mu^{\rm hvp}]^s_{\rm lat}$ up to $t_{\rm cut}$ at strange quark flavor.}\label{fig:amu_s}
  \end{center}
\end{figure}

Since our gauge ensembles have the different sea strange mass on $96^4$ and $64^4$ (see in Table~\ref{tab:lat_param}), the strange quark contribution to muon g-2 should be more significant due to its mass correction than FV effect. In Figure~\ref{fig:amu_s}, we plot both data of $[a_\mu^{\rm hvp}]^s_{\rm lat}$ and integrand of $Q^2$ integral for a comparison in strange sector. One can see that $[a_\mu^{\rm hvp}]^s_{\rm lat}$ on 96$^4$ is 6--7\% smaller than that on 64$^4$ lattice, besides, a contribution of such a discrepancy to the total muon g-2 is minor. Actually its magnitude is less than 0.5\%. 

\section{Mass correction and finite volume effect}\label{sec:mass_correction}
\begin{figure}
  \begin{center}
    \includegraphics[width=74mm]{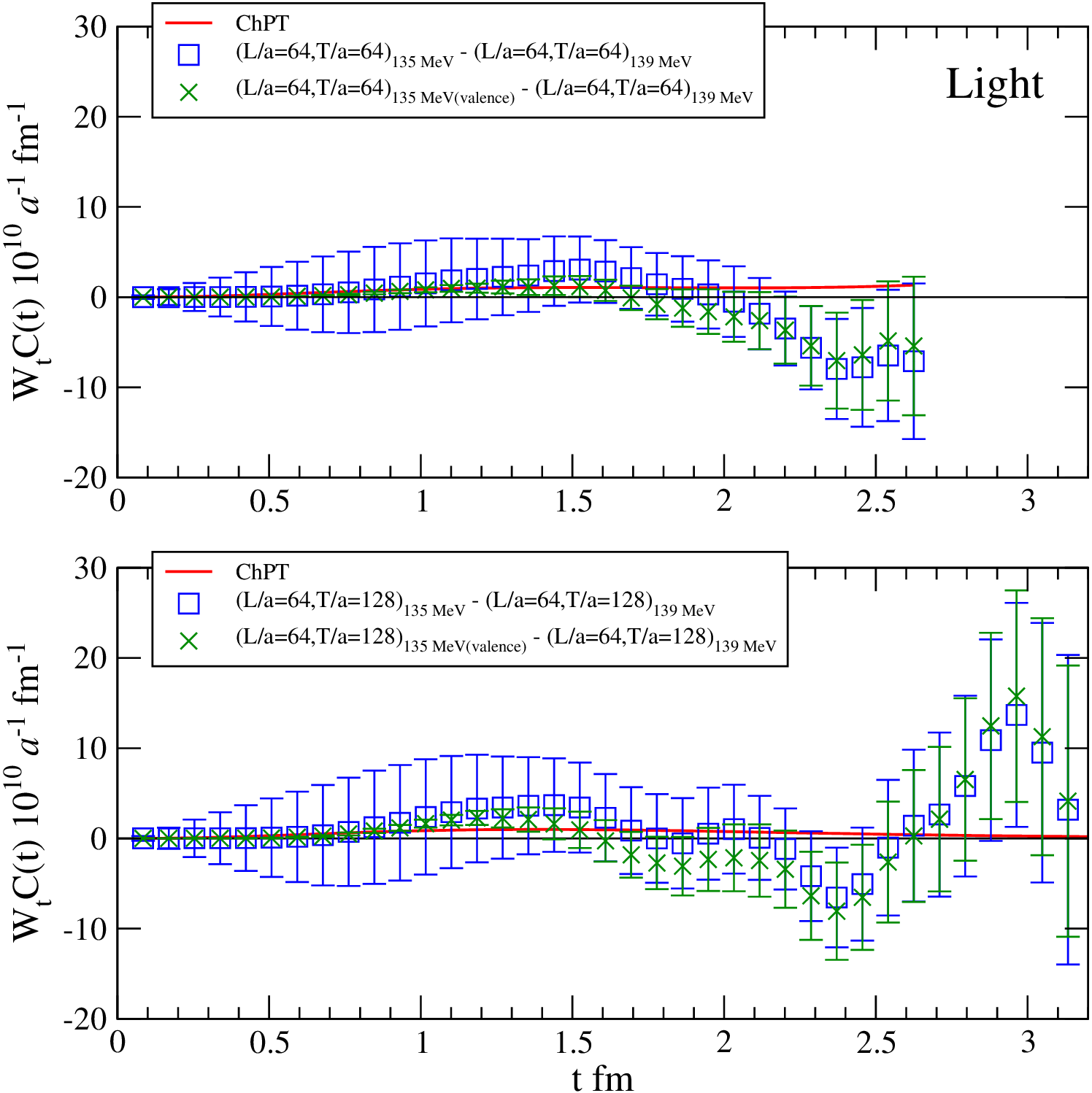}
    \hspace{5mm}
    \includegraphics[width=70mm]{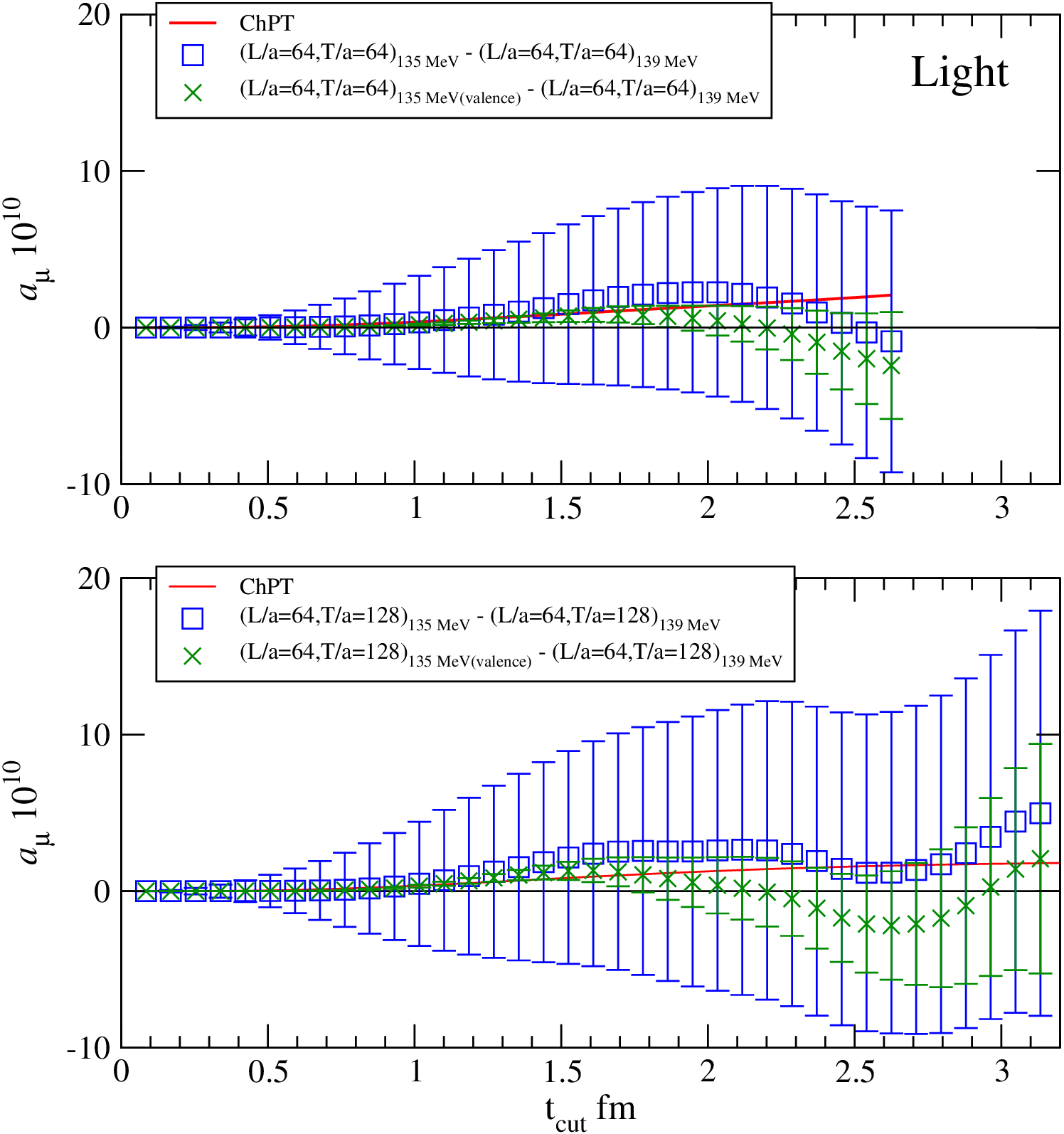}
    \caption{(Left) Difference of integrand between 135 MeV pion data using reweighting method and 139 MeV pion data on $L/a=64$. Top and bottom panels present the comparison with partially quenched case and extended temporal extension, respectively. (Right) The symbols are same as left-panel for $[a_\mu^{\rm hvp}]^l_{\rm lat}$ up to $t_{\rm cut}$. Solid (dashed) lines denote the leading order of ChPT with $T/a=64$ ($T/a=128$).}\label{fig:amu_64c_rw}
  \end{center}
\end{figure}

As mentioned before, on 96$^4$ lattice ensemble there is $\approx$7 MeV pion mass difference from 64$^4$ lattice which will be affected in the FV effect of $[a_\mu^{\rm hvp}]^l_{\rm lat}$, and so that to estimate such an effect we compare $[a_\mu^{\rm hvp}]^l_{\rm lat}$ and its integrand with shifted sea quark mass as well as valence quark by reweighting onto 135 MeV pion on 64$^4$ lattice (see those lattice parameters in Table~\ref{tab:lat_param}). As shown in Figure~\ref{fig:amu_64c_rw}, the correction of different pion mass is as slightly positive shift of $[a_\mu^{\rm hvp}]^l_{\rm lat}$ and its integrand also increases when pion mass decreases. Compared to partially quenched case, the error of reweighting factor becomes dominant contribution, especially for short time-slice, while in long time-slice its error is comparable with each other since statistical fluctuation is large in both cases. At large time-slice over $t=2$ fm on $T/a=64$, one can see that the BSP effect significantly appears in the comparison with extended temporal length $T/a=128$. The mass correction to $[a_\mu^{\rm hvp}]^l_{\rm lat}$ is evaluated as $(4\pm12)\times10^{-10}$ referred to Figure~\ref{fig:amu_64c_rw} at $t_{\rm cut}=3$ fm in $64^3\times 128$ lattice. Compared to the leading order of ChPT, such a mass correction to $[a_\mu^{\rm hvp}]^l_{\rm lat}$ is consistent with lattice data for both integrand and its time-slice summation within 1-$\sigma$ statistical error even in short time-slice. 

As presented in \cite{DellaMorte:2017dyu}, they showed the strong growth of $a_\mu^{\rm hvp}$ when $m_\pi^2$ decreases. When naively applying the linear $m_\pi^2$ behavior for their $a_\mu^{\rm hvp}$ values \footnote{In reference \cite{DellaMorte:2017dyu}, they have showed only combined results with several ansatz to perform the integral from $t_{\rm cut}$ to infinity. Since their $t_{\rm cut}$ is much smaller than ours, this estimate of mass dependence for $a_\mu^{\rm hvp}$ is naive.} between $m_\pi=0.19$ GeV and $m_\pi=0.135$ GeV, a decrease of 6\% pion mass affects roughly $7\times10^{10}$ positive contribution to $[a_\mu^{\rm hvp}]_{\rm lat}$ at light quark flavor. This value is roughly same magnitude as our estimate.

\begin{figure}
  \begin{center}
    \includegraphics[width=70mm]{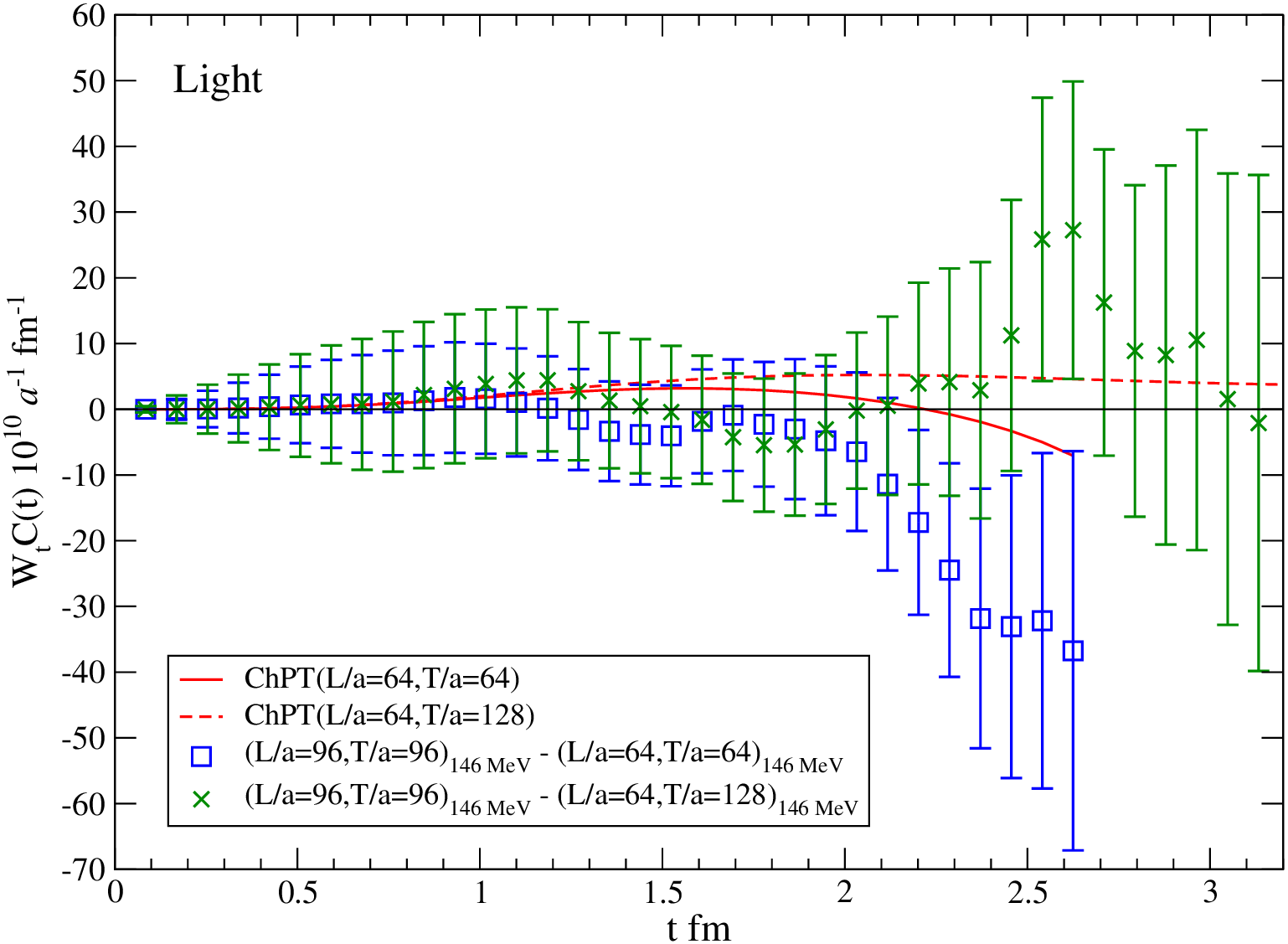}
    \hspace{5mm}
    \includegraphics[width=70mm]{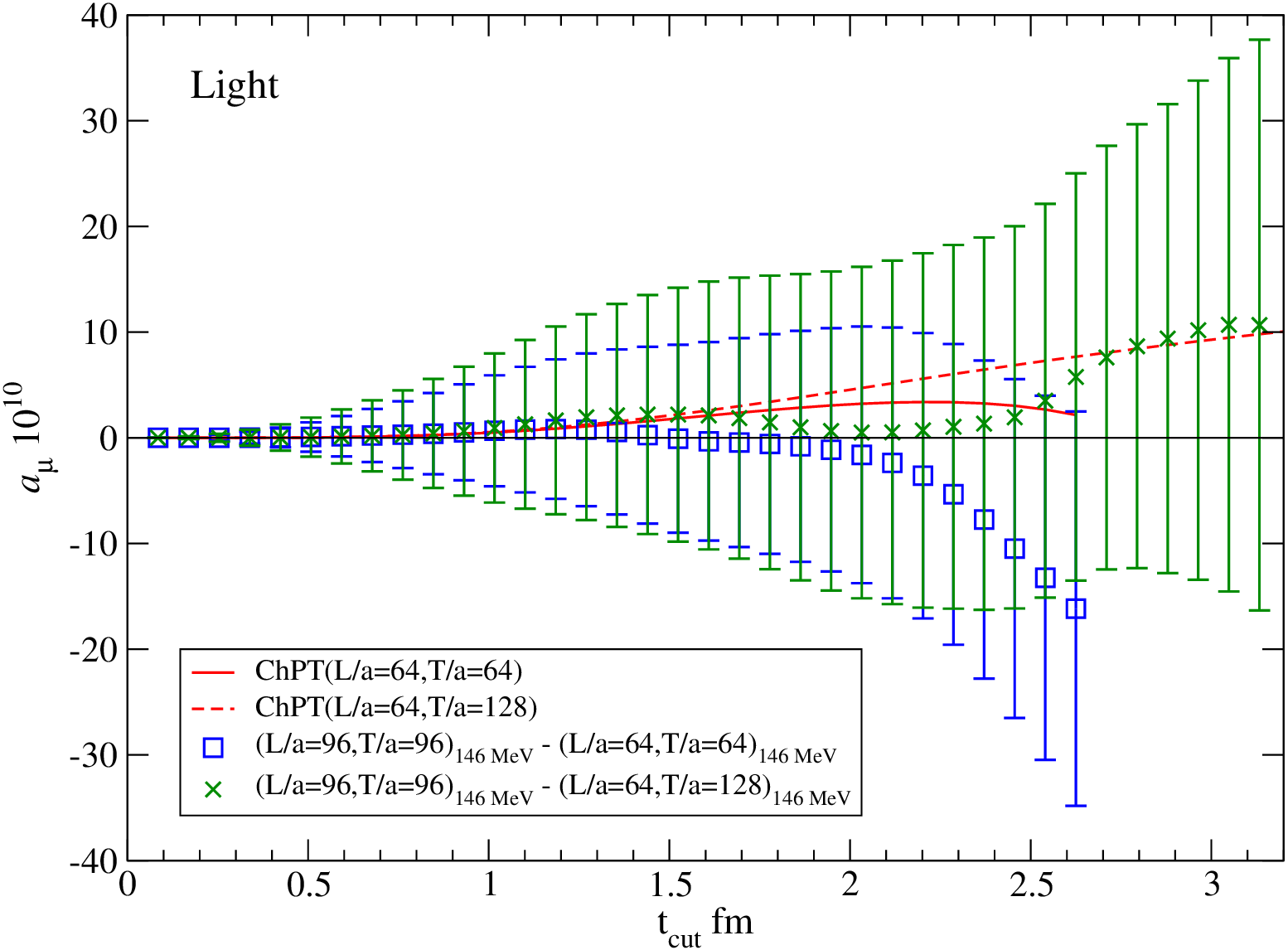}
    \caption{(Left) Finite volume effect of integrand (left) and $[a_\mu^{\rm hvp}]^l_{\rm lat}$ up to $t_{\rm cut}$ (right) with mass correction to Figure~\ref{fig:amu_diff} estimated from Figure~\ref{fig:amu_64c_rw}. Solid (dashed) lines denote the leading order of ChPT with $T/a=64^4$ ($T/a=128$).}\label{fig:amu_diff_corr}
  \end{center}
\end{figure}

To estimate FV correction between 96$^4$ and 64$^4$ at the same pion mass, we first evaluate a linearly extrapolated data into 146 MeV pion on $L/a=64$ lattice ensemble using two data of 139 MeV pion and 135 MeV pion, and we then take a difference between $L/a=96$ and $L/a=64$ in 146 MeV pion. In Figure~\ref{fig:amu_diff_corr}, we show such a comparison with different volume. One can see that the difference between data on $64^4$ lattice in 139 MeV pion and $96^4$ lattice in 146 MeV pion is canceled by the contribution of mass correction in Figure~\ref{fig:amu_64c_rw}, and as this result the FV effect is consistently zero within statistical error. Conservatively, we regard the FV correction as the volume difference at $t_{\rm cut}=$3 fm of the cross-symbol in Figure~\ref{fig:amu_diff_corr} where the BSP effect is negligible and $t_{\rm cut}$ dependence is minor. The magnitude of FV correction to $[a_\mu^{\rm hvp}]^l_{\rm lat}$ on $L=5.4$ fm is then (10$\pm$26)$\times10^{-10}$, which corresponds to $1\pm 4$\%  for the dispersive estimate $a_\mu^{\rm hvp}\approx 700\times10^{-10}$. We also plot the ChPT line in Figure~\ref{fig:amu_diff}. Although lattice estimate is still consistent with ChPT within 1-$\sigma$ statistical error, the central value on both lattice sizes is slightly over the ChPT line. To do more clear comparison, high statistics data on larger lattice size than $L=8$ fm at exactly same quark mass as $64^4$ lattice ensemble are needed, in particular for infrared region.

\section{Summary}\label{sec:summary}
This paper presents the study of finite volume (FV) correction for the connected diagram of hadronic vacuum polarization contribution to muon g-2 from the direct comparison with two volumes, $L=5.4$ fm and 8.1 fm, in purely lattice QCD, which is an independent way from the other lattice studies \cite{Chakraborty:2016mwy,Borsanyi:2017zdw,Blum:2018mom}. At the physical pion, we estimate the FV correction in time-momentum representation (TMR) method using time-slice summation of vector-vector current correlator. Using the high-statistics lattice data boosted by all-mode-averaging (AMA) method, we obtain that the light quark contribution to $[a_\mu^{\rm hvp}]_{\rm lat}$ estimated in time-slice summation on $L=5.4$ fm is $(10\pm 26)\times10^{-10}$ shift from $L=8.1$ fm as FV correction at 146 MeV pion, correspondingly $1\pm 4$\% effect to dispersive estimate of $a_\mu^{\rm hvp}\approx700\times 10^{-10}$, and $[a_\mu^{\rm hvp}]_{\rm lat}$ taken as the upper- and lower-bound in Eq.~(\ref{eq:amu_tsum_bnd}) is obtained in Eq.~(\ref{eq:amu_upper}) and (\ref{eq:amu_lower2}). In our study, compared to the estimate of the leading order of ChPT, there is no observation of significant discrepancy and it is then consistent within 1-$\sigma$ statistical error, although statistical fluctuation is still large. Here we also have a concern of the uncertainty due to truncation of integral in TMR using finite $t_{\rm cut}$, in which we expect FV correction becomes significantly large after $t_{\rm cut}$. In order to completely remove such an uncertainty, the infinite volume limit at the physcal pion is necessary to realize $[a_\mu^{\rm hvp}]_{\rm lat}$ at $t_{\rm cut}\rightarrow\infty$ in TMR. This will be done by using one more large lattice ensemble generated by PACS collaboration in the future. Our approach is also useful to check the estimate of FV correction relied on the extrapolation into physical pion and infinite volume limit simultaneously using lattice data with various pion masses and volumes~\cite{Chakraborty:2016mwy,DellaMorte:2017dyu,Giusti:2017jof,Borsanyi:2017zdw}.

Furthermore there are several systematics which has not been taken into account. First, since there is only one lattice cut-off scale on this ensemble, the lattice artifact effect involved into $a_\mu^{\rm hvp}$ can not be measured directly. We try to partially estimate it by comparison with the representation of $[\tilde a_\mu^{\rm hvp}]_{\rm lat}$ in Eq.~(\ref{eq:amu_TMR_lat2}). On such a way, lattice artifact correction appears in short time-slice, especially at $t/a=1$, and for the integral it thus affects constant shift. The magnitude of shift for $[a_\mu^{\rm hvp}]^l_{\rm lat}$ at $t_{\rm cut}=3.02$ fm is, $([\tilde a_\mu^{\rm hvp}]^l_{\rm lat}-[a_\mu^{\rm hvp}]^l_{\rm lat})\times 10^{10} = 9.11(1)$ which is roughly 2\% effect for total light flavor contribution. Note that this difference is only a consequence of discretized space-time on finite lattice spacing. The other lattice artifact caused by chiral symmetry breaking in Wilson-clover fermion should be estimated in the future using larger cut-off scale. Second, as mentioned before, this is a calculation of only connected diagram, and the disconnected piece as SU(3) flavor symmetry breaking in electromagnetic current is other missing factor in our analysis. Although several papers \cite{Blum:2015you,Borsanyi:2016lpl,DellaMorte:2017dyu} for computation of the disconnected piece in lattice QCD have reported a negative contribution to $a_\mu^{\rm hvp}$ as 1.5\%, it will be tested on this ensemble in the next work.

The future generation of several gauge ensembles with one more large volume and fine lattice spacing in PACS collaboration enables us to provide the final result in lattice QCD by simultaneously taking the infinite volume and continuum limit.

\section*{Acknowledgments}
We would like to thank PACS, RBC-UKQCD collaboration for helpful discussion and support. T.I and C.L. are supported in part by US DOE Contract DESC0012704(BNL). T.I. is also supported by JSPS KAKENHI grant numbers JP26400261, JP17H02906. C.L. is also supported by a DOE Office of Science Early Career Award. We originally develop the computation code based on Columbia Physics System(CPS) in which tuned OpenQCD system (http://luscher.web.cern.ch/luscher/openQCD/) is embeded. This work is supported in part by MEXT as ``Priority Issue on Post-K computer'' (Education of the Fundamental Laws and Evolution of the Universe) and JICFuS, and the U.S.-Japan Science and Technology Cooperation Program in High Energy Physics for FY2018. Numerical calculations were performed on the K computer in RIKEN Center for Computational Science (CCS), Hokusai at Advanced Center for Computing and Communication (ACCC) in RIKEN, XC40 at YITP in Kyoto University, BlueGene/Q in High Energy Accelerator Research Organization (KEK). The computation was also carried out using the computer facilities at Research Institute for Information Technology, Kyushu University, and conducted using the Fujitsu PRIMERGY CX600M1/CX1640M1 (Oakforest-PACS) in the Information Technology Center, The University of Tokyo. This computation is also supported by Interdisciplinary Computational Science Program No.~xg17i019, xg18i015 in Tsukuba CCS, Large Scale Simulation Program No.~16/17-26 in KEK, General use No.~G17029, G18001 at ACCC, and resources of the K computer provided by the RIKEN-CCS through the HPCI System Research project (Project ID:hp180126).

\appendix
\section{Performance test of AMA with SAP deflation}\label{appx:AMA}
In this appendix, we present the numerical test of AMA performance with two deflation methods; low-mode deflation and SAP deflation. For low-mode deflation, we compute the single-precision low-mode of Wilson-clover kernel by the Lanczos procedure with Chebychev acceleration (see in \cite{Shintani:2014vja}). On $96^4$ PACS configurations, we have 750 low-modes as $10^{-8}$ accuracy, and conjugate gradient (CG) method is used to obtain $\mathcal O^{\rm (org)}$ and $\mathcal O^{\rm (appx)}$. Its approximation having the similar correlation, to be $\Delta_r\simeq O(10^{-4})$, is generated by 600 CG iteration with low-mode projection. Using SAP deflation, the approximation is obtained with the same parameters as in Table~\ref{tab:ama_param}.

\begin{figure}
  \begin{center}
    \includegraphics[width=100mm]{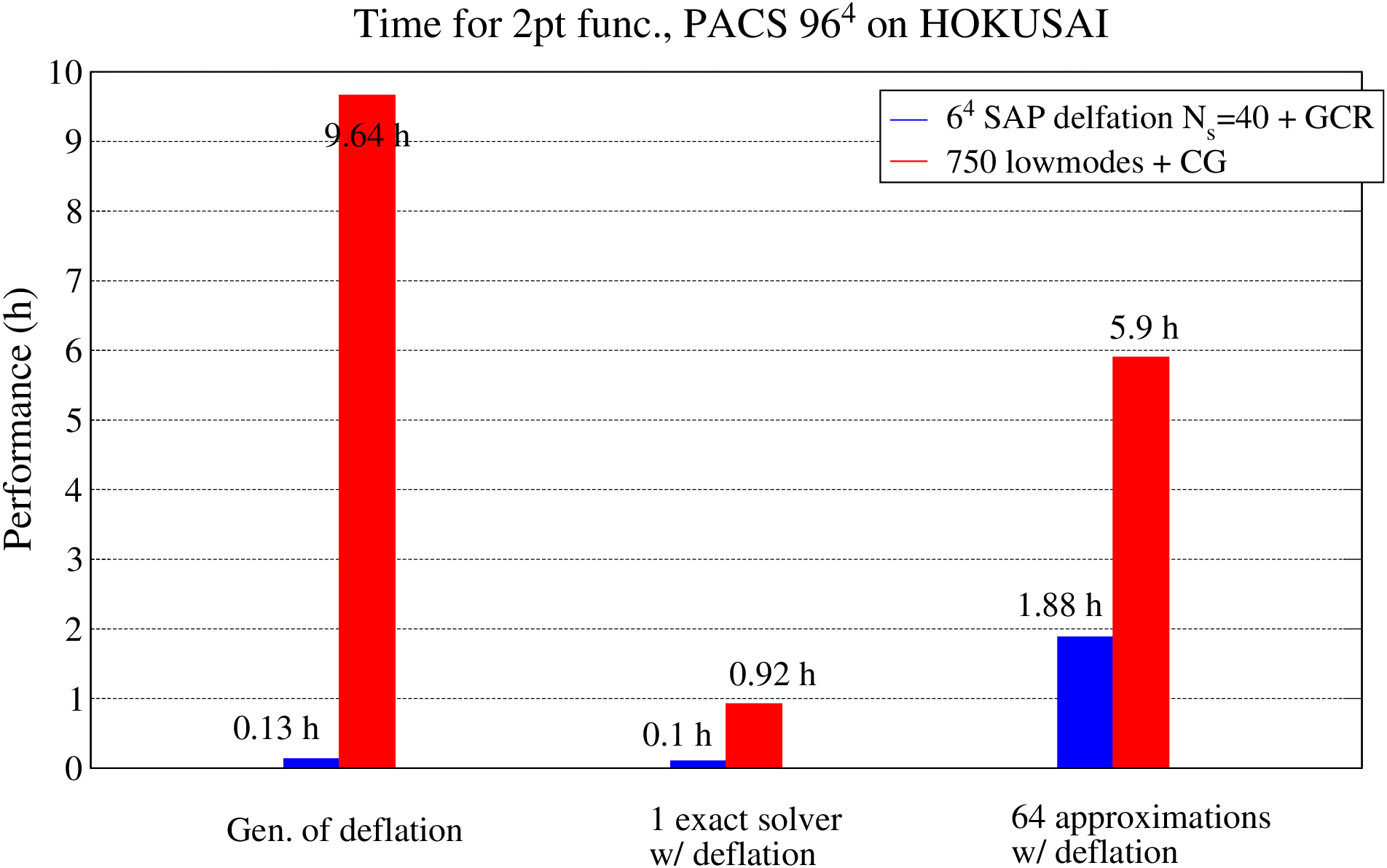}
    \caption{The comparison of elapsed time in three cases for AMA process; generation of deflation field, computation of quark propagator in exact precision $10^{-10}$ of residual norm and in the approximation. In the approximation, we compare computational time for $N_G=64$, which is the number of different source locations. We perform this test on HOKUSAI, in which the Fujitsu FX100 CPU chips having 32 cores per node are equipped. SAP deflation result is on 128 nodes, while low-mode deflation result is used as the scaled one from 256 nodes to 128 nodes assuming the strong scaling.}\label{fig:cost_sap}
  \end{center}
\end{figure}

From Figure~\ref{fig:cost_sap}, one can clearly see the computational cost for generation of deflation field is much reduced by factor 70, and also the cost of quark propagator (12 times iterative solver is performed) using GCR with SAP deflation is 9 times for exact and 3 times for approximation smaller than the case using 600 CG iteration with low-mode deflation. Totally the computational time of AMA with SAP deflation is reduced by factor 3 and more. We note that, for low-mode deflation, once we obtain low-mode vector, it enables us to recycle this data by loading from disk storage to construct the low-mode projection matrix without additional cost of the little Dirac solver during iterative process as in SAP deflation~\cite{Luscher:2007se}, however for large size of lattice, as we demonstrated in $96^4$, storing 750 eigenmodes, 6 TB disk space is needed per configuration. It turns out to be disk-consumed scheme. Furthermore, increasing the lattice size, since the number of the low-lying eigenmodes densely increase near zero, the low-mode deflation computed by the Lanczos algorithm will require a huge computation resource, for instance large memory size and disk space to store eigenmodes.

On the other hand, SAP deflation has a totally negligible cost for the generation of deflation field, and it thus does not need to store the deflation field into disk-space instead, the computation of the deflation field at each time occurs before doing the quark solver. It has an advantage to reduce the space of disk-storage. In addition, as pointed out in \cite{vonHippel:2016wid}, since SAP deflation can use $N_s$ local deflation fields by domain-decomposition of Dirac operator onto SAP block size, total memory size to store the deflation field is reduced by $O(10)$. This is also advantage to reduce the requirement of memory size.

\begin{figure}
  \begin{center}
    \includegraphics[width=100mm]{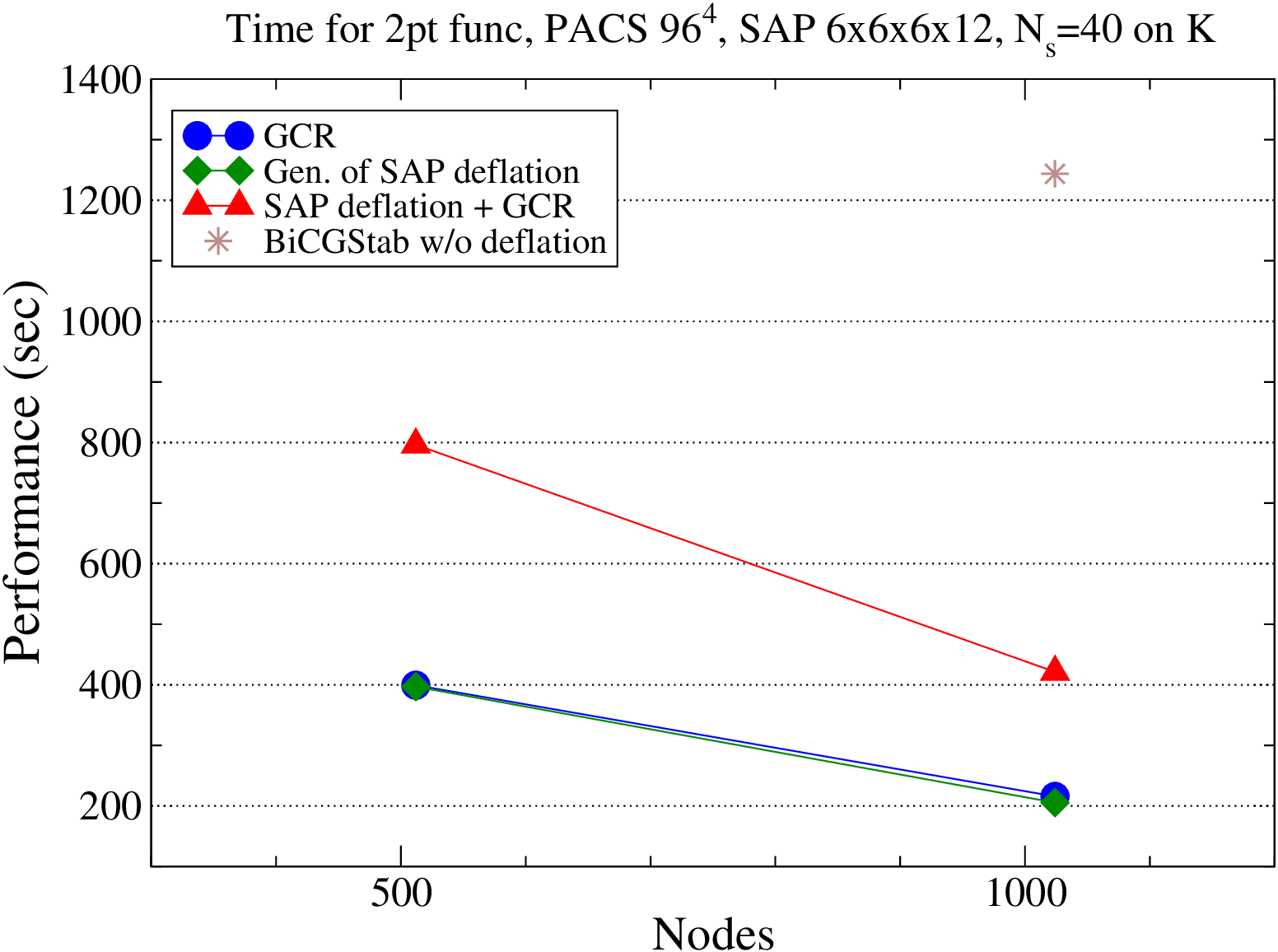}
    \caption{This plot is a strong scaling test of SAP deflation + GCR on PACS $96^4$ lattice. Here the SAP block size is $6^3\times12$ with $N_s=40$ SAP deflation fields. We carry out a computation of quark propagator on K computer, in which Fujitsu FX10 chip having 8 cores per node is equipped. Flat MPI run is used in this test. For comparison, we plot the computational time in BiCGstab method without deflation as star symbol. Note that star symbol is scaled from 2048 nodes under the assumption of strong scaling.}\label{fig:cost_scale}
  \end{center}
\end{figure}

Figure~\ref{fig:cost_scale} shows the strong scaling of SAP deflation + GCR on K computer accommodated in RIKEN-CCS. One can see the performance for both generation of SAP deflation field and GCR with deflation projection has the strong scaling from 512 nodes to 1024 nodes. We also compare the performance of SAP deflation + GCR with conventional method, which is BiCGStab solver without deflation used in \cite{Ishikawa:2015rho}. Even including the overhead to generate the deflation field, SAP deflation + GCR has more than 3 times better performance than conventional method. In the measurement, $O(10^3)$ two-point functions per configuration are needed, so that the elapsed time of quark solver is eventually dominated. Ignoring the time for generation of deflation field, GCR with SAP deflation projection can gain 6 times speed-up. Furthermore AMA can reduce such a solver time by factor 5 and more, and it thus gains more than 30 times speed-up compared to conventional one. 

\bibliographystyle{apsrev4-1}
\bibliography{ref}
\end{document}